\documentstyle[aps,prc,epsfig,manuscript]{revtex}

\newcommand{\be}{\begin{equation}}
\newcommand{\ee}{\end{equation}}
\newcommand{\bea}{\begin{eqnarray}}
\newcommand{\eea}{\end{eqnarray}}
\renewcommand{\d}{{\rm d}}
\newcommand{\PD}{{\partial}}

\begin{document}
\bibliographystyle{h-physrev}

\title{Dynamics of Hot Bulk QCD Matter: from the Quark-Gluon Plasma
to Hadronic Freeze-Out}

\author{S.A.~Bass 
  \footnote{present address:\\
    National Superconducting Cyclotron Lab., Michigan State University,
    East Lansing, MI 48824-1321  }
  \footnote{e-mail address: bass@nscl.msu.edu}}
\address{
        Department of Physics\\
        Duke University\\
        Durham, NC, 27708-0305, USA
        }

\author{A.~Dumitru
  \footnote{e-mail address: dumitru@mail-cunuke.phys.columbia.edu}}
\address{Department of Physics\\
        Columbia University\\
        538W 120th Street, New York, NY 10027, USA
        }

\maketitle

\begin{abstract}
We introduce a combined macroscopic/microscopic transport approach 
employing relativistic hydrodynamics for the early, dense, deconfined stage 
of the reaction and a microscopic non-equilibrium model for the later hadronic
stage where the equilibrium assumptions are not valid anymore.
Within this approach we study the dynamics of hot, bulk QCD matter, which
is expected to be created in ultra-relativistic  heavy ion collisions
at the SPS, the RHIC and the LHC.
Our approach is capable of self-consistently calculating the
freeze-out of the hadronic system, while
accounting for the collective flow on the
hadronization hypersurface generated by the QGP expansion.
 In particular,
we perform a detailed analysis of the reaction dynamics,
hadronic freeze-out, and transverse flow.
\end{abstract}

\pagebreak

\section{Introduction}

A major goal of colliding heavy-ions at relativistic energies is to heat up a
tiny region of space-time to temperatures as high as are thought
to have occured during the
early evolution of the universe, a few microseconds after the big
bang~\cite{KolbTurner}. In ultra-relativistic heavy-ion collisions the
four-volume of hot and dense matter,
with temperatures  above $\sim150$~MeV, is
on the order of $\sim (10$~fm$)^4$. The state of strongly
interacting matter at such high temperatures (or density of
quanta) is usually called quark-gluon plasma (QGP)~\cite{QGP}.
For a discussion of the properties and potential signatures of
such a super-dense state see~\cite{QGP,harris96a}.

A particularly interesting aspect of producing such a hot and dense
space-time
region is that QCD, the fundamental theory of strong interactions,
is expected to exhibit a transition to a new thermodynamical phase
at a critical temperature $T_C\sim100-300$~MeV.
This phase transition has been observed in
numerical studies of the thermodynamics of QCD at vanishing
net baryon charge on lattices~\cite{LQCD}.
It is the only phase transition of a fundamental theory that is
accessible to experiments under controlled laboratory conditions.

In this paper we shall investigate the dynamics of relativistic
heavy ion collisions within a novel transport approach combining
a macroscopic and a microscopic model.
We shall focus here on collision systems currently 
under investigation at the CERN Super-Proton-Synchrotron (SPS), the
Relativistic Heavy Ion Collider (RHIC) at BNL and the future
Large Hadron Collider (LHC) at CERN.

We shall work in natural units $\hbar=c=k=1$ throughout the paper.

\section{General Aspects of Matching Fluid Dynamics to
Microscopic Transport}

In this section we discuss general aspects and assumptions of our
model for the space-time evolution of high-energy heavy-ion reactions.
In particular, we introduce fluid dynamics for the early, hot
stage, and the matching to microscopic transport for the later,
more dilute stages of the reaction. Within this section, quantities
without subscript refer to the fluid, while properties of the
microscopic transport theory carry the subscript $micro$.

\subsection{Transport Equation for Incoherent Quanta / Particles}
The most basic assumption of our model for the evolution of high-energy
heavy-ion reactions is that 
at the initial time\footnote{Our choice of space-time
variables is described in more detail below; for the moment, we
assume that suitable variables have been chosen, and that
the hypersurfaces of homogeneity are time-orthogonal everywhere.}
$t=t_i$ the highly excited space-time domain
produced in the impact can be viewed as being populated by incoherent
quanta on the mass-shell.
Thus, the system can be described by a distribution function
$f_i(x^\mu,p^\nu)$, where $x^0=t_i$, $p^0=\sqrt{\vec{p}^2+m_i^2}$,
and $i$ labels different species of quanta.
We will not discuss here how such a state of high entropy density could
possibly be reached. That discussion is out of the scope of the present
manuscript. Our work addresses the subsequent evolution of that initial state
up to the so-called freeze-out of strong interactions in the system.

The semi-classical evolution of the distribution function in the forward
light-cone is described by means of a so-called {\em transport} equation,
e.g.\ the Boltzmann equation~\cite{ChVW}
\be \label{Beq}
p\cdot\partial f_i(x^\mu,p^\nu) = {\cal C}_i\quad.
\ee
${\cal C}_i$ is the collision kernel, describing gain or loss
of quanta (particles) of species $i$ in the phase-space cell around
$(x^\mu,p^\nu)$
due to {\em collisions}. Note that we have dropped possible
classical background fields in eq.~(\ref{Beq}).

\subsection{Moments of the Transport Equation: Hydrodynamics for the
hottest stage}
For the problem at hand, however, the usefulness of eq.~(\ref{Beq})
is rather limited. A major difficulty is that to obtain an analytical
or numerical solution, in most cases one
has to introduce an expansion of the collision kernel in terms of
the number of incoming particles per ``elementary'' collision~\cite{ChVW}
(In most practical applications that expansion is even truncated at
the level of binary collisions, $2\rightarrow n$). Obviously, the
expansion is ill-defined at very high densities.
A second major problem is to describe the hadronization process,
i.e.\ the dynamical conversion of quarks and gluons into hadrons,
on the microscopic level. Several interesting approaches to describe the
hadronization of a plasma of quarks and gluons microscopically
have been proposed in the literature, cf.\ e.g.~\cite{microhad} and references
therein.
However, due to the very complicated nature of this process, many of those
models have to involve some kind of ad-hoc prescriptions, which have
quite significant impact on the results.
A first-order QCD phase transition, as assumed in the following,
is particularly difficult to model microscopically.

At present we are not able to solve these problems in a fully
satisfactory way. We can, however, circumvent them to some extent
if we are mainly interested in the bulk dynamics of
a hot QCD system. In this case we can employ relativistic ideal
hydrodynamics~\cite{LLH}
for the very dense stage of the reaction up to hadronization.

Let us thus assume that it is feasible to employ the
continuum limit. The first two moments of eq.~(\ref{Beq})
yield the continuity equations for the conserved currents
and for energy and momentum~\cite{ChVW},
\be \label{conteq}
\partial\cdot N_i=0\quad,\quad\partial\cdot \Theta=0\quad.
\ee
In the following, we will explicitly consider only one
conserved current, namely the (net) baryon current. All other currents,
as e.g.\ strangeness, charm, electric charge etc.,
will be assumed to vanish identically (due to local charge
neutrality and an ideal fluid) such that the corresponding
continuity equations are trivially satisfied.

Ideal fluid dynamics goes even further and assumes that the momentum-space
distributions in the local
rest frame are given by either Fermi-Dirac or Bose-Einstein
distribution functions, respectively. Dissipation and heat
conduction which arise from higher moments are neglected.
Since we restrict fluid-dynamics to the high-temperature and
high-density stage, this
approximation is at least logically consistent.
In future work, it will be important though to check its
quantitative accuracy.

The density of secondary partons in the central region of
high-energy nuclear collisions is very high. According to
present knowledge, it is likely that the central region evolves from a stage
of pre-equilibrium towards a QGP in local thermal
equilibrium~\cite{geiger92D46,Biro:1993qt,Wong:1996va,Gyulassy:1997ib},
despite the large expansion rate.
On the other hand, the very same calculations do not seem to support rapid
chemical equilibration (in particular of the quarks), cf.\ also
\cite{GeigKap,Srivastava:1996qd,Elliott:1999uz}. However,
in most publications interactions among the secondary partons (and in
particular particle production via inelastic processes) were treated
perturbatively. Since the running coupling in a thermal plasma
with $T\le1$~GeV is not very small, one can not exclude sizeable
contributions from processes involving higher powers of
$\alpha_s$~\cite{Wong:1996va}. Moreover, in addition to the semi-hard
partons there might exist a coherent color field in the central
region (between the receding nuclear ``pancakes'') which
produces additional quark-antiquark pairs in its decay~\cite{cohstr}.
In any case, we will not argue in favor nor against rapid $q\overline{q}$
production and chemical equilibration but simply assume that the quark
densities at the initial time of the hydrodynamical expansion are close to
their chemical equilibrium values.
At least for Pb+Pb at CERN-SPS energy, where experimental data already
exists, this is basically the only way for our model to account for the
fact that {\em measured} hadron multiplicity ratios are close to their
chemical equilibrium values~\cite{bms}.
Since the expansion rate after hadronization is too large for
``chemical cooking'' (and in particular for strangeness equilibration),
as will be discussed in section~\ref{ChemFO},
it would be virtually impossible to achieve approximate chemical equilibrium
during the later hadronic stages if starting from a QGP far off chemical
equilibrium, cf.\ also~\cite{Stock:1999hm}.

The general picture as described above is summarized in the space-time
diagram depicted in Fig.~\ref{spt_diagram}. We assume that ideal fluid
dynamics is a reasonable approximation between the ``initial'' time
$t_i$ and the hadronization hypersurface. After that, we will switch
to a microscopic description employing the binary collision
approximation for the collision kernel. In particular, we will employ
the Ultrarelativistic  Quantum Molecular Dynamics (UrQMD)
transport model, see below.

\subsection{Microscopic Transport from Hadronization to Freeze-Out}
One may ask why it is not sufficient to rely on hydrodynamics up to
some rather late stage of the reaction, after which one postulates
that all particle momenta are ``frozen'' and thus are equal to those measured
in the detector at $t_f\rightarrow\infty$. That approach has been
applied to nuclear collisions by many authors, for recent work cf.\
e.g.~\cite{BSch,JSoll,DumKat,DumRi}, and leads to reasonable
results for the single-particle spectra of the most
abundant hadron species $\pi$, $K$, $p$, $\Lambda$.
However, the following limitations arise:

First, the evolution must clearly be non-ideal in the late stages of the
reaction~\cite{relax},
as the system approaches ``freeze-out''. This can manifest
in decoupling of various components of the fluid (e.g.\ pions and nucleons),
i.e.\ each component develops an individual collective velocity~\cite{LevBM}.
Another aspect is that the $-p\d V$ expansion work performed by the fluid
can be partly compensated by entropy production ($+T\d S$) such that
the expansion may even become isoergic, $\d E=0$, instead of isentropic,
$\d S=0$~\cite{Danie}.

Moreover, since each hadron
is propagated individually, and its interactions with other hadrons
are described on the basis of elementary processes, microscopic transport
models offer the opportunity to {\em calculate} the freeze-out conditions
instead of just putting them in by hand as is done in the purely
fluid-dynamical approaches~\cite{BSch,JSoll,DumKat,DumRi}.
There one assumes that freeze-out occurs whenever some criterion
is fulfilled, e.g.\ when the temperature drops below some ``guessed'' value.
In contrast, the non-truncated transport eq.~(\ref{Beq}) can describe
self-consistently the freeze-out of the system: no decoupling hypersurface
is imposed by hand, but rather is determined by an interplay between
the (local) expansion scalar $\partial\cdot
u$~\cite{KolbTurner,locexp,hung98a,ad99}
(where $u$ is the four-velocity of the local rest-frame),
the relevant elementary
cross sections and decay rates, and the equation of state (EoS), which
actually changes dynamically as more and more hadron species decouple.
This is obviously a key point for being able to study and predict the
dependence of the final state on collision energy (i.e.\ on the initial
entropy or energy density), system size etc., instead of just fitting
it by an appropriate choice of a freeze-out hypersurface.
Note e.g.\ that the nucleons emerging from the QCD hadronization phase
transition in the early universe were able to maintain chemical
equilibrium down to temperatures of about
$\sim50$~MeV \cite{KolbTurner}. In heavy-ion collisions at CERN-SPS
energies, however,
one finds chemical freeze-out temperatures on the order of $140-
160$~MeV \cite{bms}.
The origin of this difference lies in the much smaller expansion
rate (Hubble constant) of the early universe as compared to a high-energy
heavy ion collision~\cite{ad99}, and can only be explained
within kinetic theory but not within pure hydrodynamics.

Another complication arises from the fact that close to the freeze-out
hypersurface the freeze-out process
feeds back on the evolution of the fluid~\cite{bugaev,cs_fo,feedback}. 
This will in general deform the freeze-out hypersurface, say an isotherm
of given temperature $T_{fo}$. It will differ from that found
a posteriori from the solution of
eqs.~(\ref{conteq}) in the whole forward light-cone.
Furthermore, the idealization
that the transition from ideal flow to free-streaming occurs on a
sharp hypersurface, i.e.\ a three-volume in space-time, is rather crude.
One instead expects a smooth transition as the temperature (and the
density of particles) decreases, cf.\ e.g.\ the discussion
in~\cite{Grassi97}.
This is supported by studies of the hadron kinetics
close to freeze-out with realistic cross-sections~\cite{microFO,bassdum99},
cf.\ also section~\ref{FO_hupersurf}.

Finally, it is likely that the freeze-out is not universal for all
hadron species, simply because their transport cross-sections are very
different. One can therefore hardly assume that all hadron species decouple
on the same hypersurface~\cite{hung98a,microFO,bassdum99,kinDec}.
The clearest example for this is the transverse momentum distribution
of $\Omega$ baryons obtained by the WA97 collaboration~\cite{expOm}
for Pb+Pb collisions at CERN-SPS energy, $\sqrt{s}=17A$~GeV.
Unlike is the case for pions, nucleons, anti-nucleons, and lambdas,
the $p_T$ distribution of omegas as calculated within hydrodynamics with
freeze-out on the $T=T_{fo}=130$~MeV hypersurface~\cite{DumRi} is
much stiffer than the experimental finding. Indeed, more
detailed kinetic treatments which explicitly account for the small
transport cross-section of $\Omega$ baryons in a meson-rich hadron gas,
emerging either from fragmentation of longitudinally stretched
color-strings~\cite{thOm} or an incoherent hot plasma of quarks and
gluons~\cite{dumitru99a}, show that these multiple-strange particles
freeze out earlier and pick up less collective transverse flow than
pions and nucleons, for example.

\subsection{Transition from Fluid Dynamics to Microscopic Transport}
\label{FD_MT_general}
A few remarks on the transition from hydrodynamics to microscopic
transport are in order here. In general, one
should introduce source terms $-\partial\cdot N_{micro}$ and
$-\partial\cdot \Theta_{micro}$ on the right-hand-sides of eqs.~(\ref{conteq}),
where 
\bea \label{Nkinth}
N_{micro}^\mu(x) &=& \sum_i\int \frac{\d^3k}{k^0_i} k^\mu
f_{i,{\rm micro}}(x,k)\quad, \\
\Theta_{micro}^{\mu\nu}(x) &=& \sum_i\int \frac{\d^3k}{k^0_i}
 k^\mu k^\nu f_{i,{\rm micro}}(x,k)
\label{Tkinth}
\eea
denote the net baryon current and the energy-momentum tensor of
the microscopic transport model, respectively. Accordingly,
external sources of particles have to be introduced in the
transport equation, which model the net-baryon charge and
energy-momentum transfer from the fluid. This way a self-consistent
solution in the whole forward light-cone, starting from the
initial hypersurface $t=t_i$ could be obtained.

\begin{figure}[htp]
\centerline{\epsfig{figure=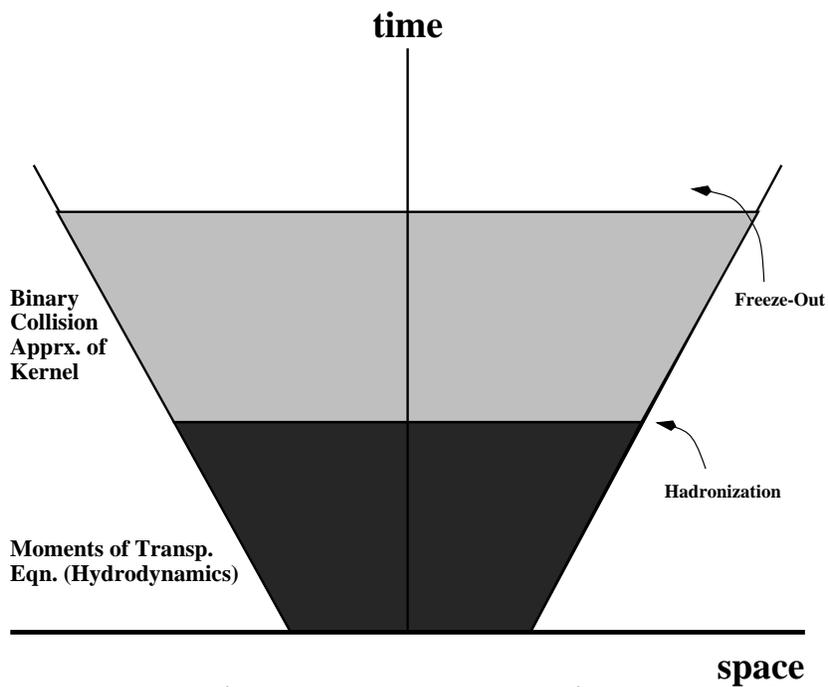,width=3.5in,angle=-90}}
\caption{Schematic overview of the space-time evolution of a
high-energy heavy-ion collision as assumed in the model presented
here.}
\label{spt_diagram}
\end{figure}

However, if a space-time region bounded by a hypersurface
$\sigma^\mu_H$ exists where fluid-dynamics is
an adequate approximation, one can choose an arbitrary hypersurface
$\sigma^\mu_{switch}$ within this region
where to switch from eqs.~(\ref{conteq}) to~(\ref{Beq}). One can
then simply assume $N_{micro}\equiv0$ and $\Theta_{micro}\equiv0$
in the interior\footnote{``Interior'' meaning towards the origin of our
space-time diagram, Fig.~\ref{spt_diagram}.}, and $N\equiv0$,
$\Theta\equiv0$ in the exterior.
On that hypersurface, one sets $N_{micro}\equiv N$,
$\Theta_{micro}\equiv \Theta$. This is because hydrodynamics is a
limiting case of eq.~(\ref{Beq}), and this more general
transport equation will automatically recover the fluid-dynamical
solution in the space-time region between $\sigma^\mu_{switch}$
and $\sigma^\mu_H$.

For the particular model discussed here,
$\sigma^\mu_{switch}$ can not precede the hadronization hypersurface
since our microscopic transport model deals with color-singlet
states, only. Also, it employs the binary collision approximation
of the kernel, which becomes less justified in the hot and dense
stage preceding hadronization.

Furthermore, as will be discussed in more detail below,
in high-energy heavy-ion collisions it turns out that
the boundary of validity of (ideal) fluid dynamics, $\sigma^\mu_H$,
can not extend far into the post-hadronization stage (the less the
higher the collision energy).
Thus, we conclude that the hadronization hypersurface is the most
natural choice for the switch from eqs.~(\ref{conteq}) to~(\ref{Beq}).

The phase-space distribution of particles of species $i$ on
$\sigma^\mu_{switch}$ is then given by~\cite{CF}
\be \label{CF_gen}
E_i \frac{\d N_i}{\d^3p} = \int \d\sigma\cdot p \, \, f(p\cdot u),
\ee
where $u^\mu$ is the four-velocity of the local rest-frame.
An explicit expression for the geometry suitable for high-energy
collisions will be given below. For time-orthogonal hypersurfaces
as depicted in Fig.~\ref{spt_diagram} one has
\be
\d\sigma_\mu\Big|_{t={\rm const.}}=(\d^3 x,\vec{0})\quad,
\ee
the left-hand-side of eq.~(\ref{CF_gen}) being simply $E_i f_{i,micro}$.

It is clear that by construction the microscopic transport starts
from a state of local equilibrium on $\sigma^\mu_{switch}$, the hypersurface
where the switch is performed. The local energy density,
net baryon density, and
collective expansion velocity are those obtained from the
hydrodynamical solution. Thus, the conserved currents and the
energy-momentum tensor of the microscopic transport theory
assume the form appropriate for ideal fluids~\cite{LLH},
\bea
N_{micro}^\mu &=& \rho u^\mu\quad,\\
\Theta_{micro}^{\mu\nu}
&=& \left( \epsilon +p_{micro}\right) u^\mu u^\nu
 -p_{micro} g^{\mu\nu} \quad.
\eea
Now, in order that
$\Theta_{micro} = \Theta$ on $\sigma^\mu_{switch}$, the pressure at given
energy and baryon density must equal that of the fluid-dynamical
model, i.e., the equations of state in local thermodynamical equilibrium
must be the same. In general this requirement is non-trivial. For ideal gases,
however, it can be obeyed by simply including the same states $i$ in the
microscopic transport ~(\ref{Beq}) as in the grand partition function
which is used to calculate the equation of state employed in hydrodynamics.
We shall discuss this point in more detail when presenting our
specific equation of state below.

\begin{figure}[htp]
\centerline{\epsfig{figure=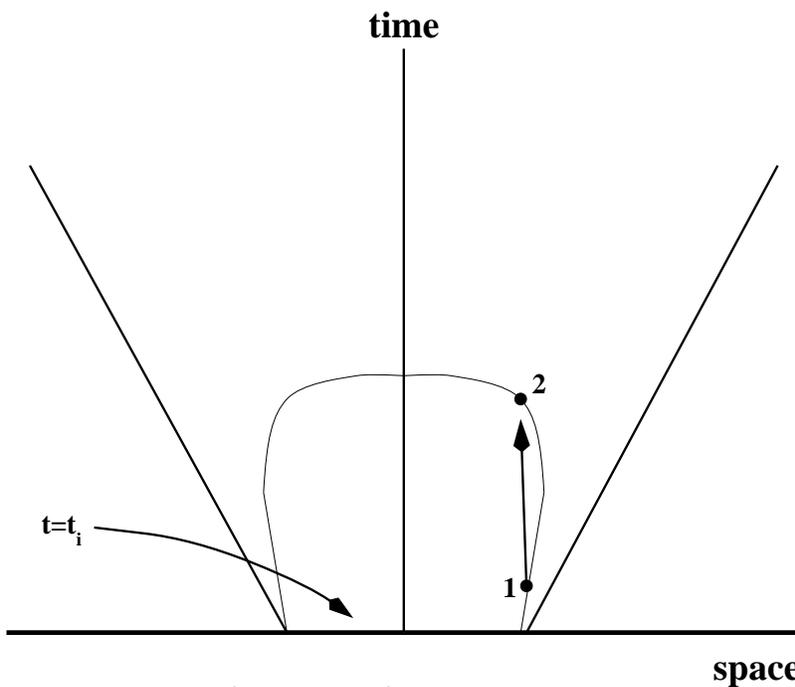,width=3.5in,angle=-90}}
\caption{Schematic example of a hypersurface with both space-like
and time-like parts.}
\label{tl_hs}
\end{figure}

We finally briefly discuss one last aspect of the switch from
fluid dynamics to microscopic transport on some hypersurface
$\sigma^\mu_{switch}$. As already mentioned above, this hypersurface is
assumed to be {\em within} the region of validity of ideal hydrodynamics,
and should be identified with the hadronization
hypersurface. However, the latter will in general also exhibit
time-like parts (points where the normal vector on $\sigma^\mu_{switch}$ is
space-like). A schematic example is given in Fig.~\ref{tl_hs}.
The initial condition of the microscopic transport on
$\sigma^\mu_{switch}$ can now {\em not} be chosen arbitrarily.
This is clear from the fact that the points 1 and 2, for example,
are causally connected.
The simplest way to prevent violation of the evolution equations is to
specify initial conditions on a purely space-like hypersurface (e.g.\
$t=t_i$) and to {\em employ} the dynamical equations, in our case the
continuity equations~(\ref{conteq}), to calculate $\Theta$ and $N$
on $\sigma^\mu_{switch}$. This way the states of the system at 1 and
at 2 are consistent.

However, one problem with switching on a hypersurface with time-like parts
remains. As discussed above, eq.~(\ref{CF_gen}) conserves energy-momentum
and (net) baryon charge. For this to hold, it actually does not count
the flow of the currents from the inside to the outside of
$\sigma^\mu_{switch}$ but, actually, the {\em net} flow. That is,
the difference of outflowing and inflowing charge, momentum etc.
The inflow is due to those parts of the thermal distribution
function $f(x,p)$ which move into the opposite direction than
the fluid. Due to the exponential tails of $f$ such particles
clearly always exist, but their number decreases strongly if the
collective flow is strong. In this case the locally isotropic
momentum-space distribution is strongly boosted.

Thus, the in-current is obtained under the assumption that
within an infinitesimal region on {\em both} sides of
the hypersurface there is hydrodynamic flow and local thermodynamical
equilibrium. For this reason, $\sigma^\mu_{switch}$ must be entirely
within the region of validity of fluid dynamics, $\sigma^\mu_H$.
Again, in this case eq.~(\ref{CF_gen}) gives the {\em net}
flow of all the currents from the fluid-region to the
region where we apply the microscopic transport.

The problem is, however, that in some part of momentum and coordinate
space the left-hand-side of eq.~(\ref{CF_gen}) can be negative.
This means that the ingoing flow exceeds the outgoing flow.
These ``negative contributions'' were already discussed by
several authors~\cite{bugaev,cs_fo,bernardri}. Since we will
interpret $\d^3p\,\d\sigma\cdot p f(p\cdot u)/E$ as a probability
distribution we have to require positive definiteness. This can
either be achieved by multiplying with a cut-off function
$\Theta(\d\sigma\cdot p)$, which leads to a slight violation
of the conservation laws; or by integration over sufficiently
large bins in momentum and coordinate-space, and random redistribution
of the particles within the bins, which smears out the
distribution over momentum and coordinate space.

A rigorous solution of this problem requires to
introduce the above-mentioned source-terms in the fluid-dynamical
evolution equations as well as in the microscopic transport
equation. However, for the cases studied here the negative
contributions were not relevant. The main reason is that
the collective flow velocity on the time-like parts of the
hypersurface is close to one,
such that net flow of particles from the microscopic transport
to hydrodynamics does not occur. For non-relativistic
flow on $\sigma^\mu_{switch}$, however, the negative contributions
would be more serious.

\section{Specific Model for High-Energy Heavy-Ion Collisions}
In the present paper we shall use hydrodynamics to model a
first order phase transition from a QGP to a hadronic fluid,
and combine it with a microscopic
transport calculation for the later, purely hadronic stages of the reaction.

In the following sections 
we describe the particular hydrodynamical and transport
models employed here, cf.\ also~\cite{DumRi} and~\cite{uqmdref1}.

\subsection{Scaling Hydrodynamics}
As already mentioned above, hydrodynamics for hadronic collisions
is defined by (local)
energy-momentum and net baryon charge conservation,
\be \label{Hydro}
\partial\cdot \Theta=0 \quad,\quad
\partial\cdot N_B=0\quad.
\ee
$\Theta^{\mu\nu}$ denotes the energy-momentum tensor, and $N_B^{\mu}$ the
current of net baryon charge.

For ideal fluids, the energy-momentum tensor and the net baryon current
assume the simple form~\cite{LLH}
\be \label{idfluid}
\Theta^{\mu\nu}=\left(\epsilon+p\right) u^\mu u^\nu -p g^{\mu\nu}
\quad,\quad
N_B^\mu=\rho_B u^\mu \quad,
\ee
where $\epsilon$, $p$, $\rho_B$ are energy density, pressure, and net baryon
density in the local rest frame of the fluid, which is defined by
$N_B^\mu=(\rho_B,\vec{0})$. 
Let us, in the following, work in the metric
$g^{\mu\nu}={\rm diag}(+,-,-,-)$.
$u^\mu=\gamma(1,\vec{v})$ is the four-velocity of the fluid
($\vec{v}$ is the three-velocity and $\gamma=(1-\vec{v}^2)^{-1/2}$ the
Lorentz factor). The system of partial differential
equations~(\ref{Hydro}) is closed by choosing an equation of state (EoS)
in the form $p=p(\epsilon,\rho_B)$, cf.\ below.

For simplicity, we assume a cylindrically symmetric transverse expansion 
with a longitudinal scaling flow profile, $v_z=z/t$~\cite{Bj}.
At $z=0$, equations~(\ref{Hydro}) reduce to
\bea \label{2Deq}
\partial_t E+\partial_T \left[\left(E+p\right)v_T\right] &=&
  - \left(\frac{v_T}{r_T}+\frac{1}{t}\right)\left(E+p\right)\quad,\nonumber\\
\partial_t M+\partial_T \left(Mv_T+p\right) &=&
  - \left(\frac{v_T}{r_T}+\frac{1}{t}\right) M\quad,\\
\partial_t R+\partial_T \left(Rv_T\right) &=&
  - \left(\frac{v_T}{r_T}+\frac{1}{t}\right) R\quad, \nonumber
\eea
where we defined $E\equiv \Theta^{00}$, $M\equiv \Theta^{0T}$, 
and $R\equiv N_B^0$.
In the above expressions, the index $T$ refers to the transverse component
of the corresponding quantity.

The set of equations~(\ref{2Deq}) describes the evolution in the $z=0$
plane. Due to the assumption of longitudinal scaling,
the solution at any other $z\neq0$ can be simply obtained by a
Lorentz boost. The above equations also imply
\be \label{peta}
\left.\frac{\partial p}{\partial\eta}\right|_{\tau,r_T}=0 \quad,
\ee
where $\eta\equiv{\rm Artanh}~v_z$ and
$\tau\equiv\sqrt{t^2-z^2}$. This means that on $\tau={\rm const.}$
hypersurfaces
pressure gradients in rapidity direction vanish, and there is no flow
between adjacent infinitesimal rapidity slices. However, only for net
baryon free matter, $\rho_B\equiv0$, does this automatically also mean
that the temperature $T$ is independent of the longitudinal fluid rapidity
$\eta$. In the case $\rho_B\neq0$ equation~(\ref{peta}) only demands
\be \label{Tmueta}
\left. s\frac{\partial T}{\partial\eta}\right|_\tau +
\left. \rho_B\frac{\partial \mu_B}{\partial\eta}\right|_\tau=0\quad.
\ee
$s$ and $\mu_B$ denote entropy density and baryon-chemical potential,
respectively. If other charges like strangeness or electric charge are locally
non-vanishing, additional terms appear.
Equation (\ref{Tmueta}) does not imply that the rapidity
distribution of produced particles is flat (i.e.\ independent of rapidity)
or that the rapidity distributions of various species of hadrons, e.g.\
pions, kaons, and nucleons, are similar. {\em Any rapidity-dependent}
$T$ and $\mu_B$ that satisfy eq.~(\ref{Tmueta}) are in agreement with
energy-momentum and net baryon number conservation, as well as with
longitudinal scaling flow $v_z=z/t$~\cite{Dumitru:1995bu}.
Note also that non-trivial solutions of eq.~(\ref{Tmueta}) in general also
yield $\PD\mu_S/\PD\eta\ne0$ on the hadronization hypersurface (i.e.\ a
rapidity-dependent strangeness-chemical potential), even if
the strangeness-density $\rho_S=0$ everywhere in the forward light-cone.
In this paper, however, we do not
explore the rapidity dependence of the particle spectra, and thus simply
assume that $T$ and $\mu_B$ are independent of $\eta$.

The fluid-dynamical evolution equations can be solved numerically on
a discretized space-time grid, cf.\ e.g.~\cite{RiBe,RiGy2}.

\subsection{Equation of State}
To close the system of coupled equations of hydrodynamics, an equation of
state (EoS) has to be specified. 
From eq.~(\ref{idfluid}) it follows that for an ideal gas the pressure $p$ is
given by
\be
p = q\cdot\left(q\cdot \Theta\right)\quad,
\ee
where $q^\mu$ is orthogonal to $u^\mu$ and normalized to $q\cdot q=-1$.
In particular, in the local rest-frame $u^\mu=(1,\vec{0})$, we
can choose $q^\mu\propto(0,1,0,0)+(0,0,1,0)+(0,0,0,1)$. Then, from
the definition of the energy-momentum tensor from kinetic theory,
eq.~(\ref{Tkinth}), we obtain
\be \label{p_idgas}
p(T,\mu_B,\mu_S) =
 \sum_i\int \frac{\d^3k}{k^0_i} \frac{\vec{k}^2}{3} f_i(k;T,\mu_B,\mu_S)
\quad.
\ee
The sum over $i$ extends over the various particle species. The grand
canonical potential is given by $\Omega=-pV$, where $V\equiv\int
\d\sigma\cdot u$ denotes the three-volume of the given hypersurface of
homogeneity. All other quantities can be obtained via standard thermodynamical
relationships. E.g., the densities of entropy, net baryon charge, and
energy are given by
\bea
s(T,\mu_B,\mu_S) &=& \frac{\partial p(T,\mu_B,\mu_S)}{\partial T} 
\quad, \label{s_idgas}\\
\rho_B(T,\mu_B,\mu_S) &=& \frac{\partial p(T,\mu_B,\mu_S)}{\partial \mu_B}
\quad, \label{rho_idgas}\\
\rho_S(T,\mu_B,\mu_S) &=& \frac{\partial p(T,\mu_B,\mu_S)}{\partial \mu_S}
\stackrel{!}{=}0\quad, \label{rhos_idgas}\\
\epsilon(T,\mu_B,\mu_S) &=& Ts-p+\mu_B\rho_B\quad. \label{e_idgas}
\eea
From $p(T,\mu_B,\mu_S)$, $\rho_B(T,\mu_B,\mu_S)$ and
$\epsilon(T,\mu_B,\mu_S)$ one can
construct the function $p(\epsilon,\rho_B)$ which is needed to close
the system of continuity equations~(\ref{Hydro}).

So far, we discussed an ideal gas, only. However,
lattice QCD predicts a phase transition
from ordinary nuclear matter to a so-called quark-gluon plasma (QGP) at
a critical temperature of $T_C=140-160$~MeV~\cite{LQCD} (for $\rho_B=0$).
We will employ a very simple and intuitive, though not very well
justified description of this phase transition.
We model the high-temperature phase as an ideal gas of $u$, $d$,
$s$ quarks (with masses $m_u=m_d=0$, $m_s=150$~MeV), and gluons, employing
the well-known MIT bag model EoS~\cite{QGP,MIT}. In this model the
non-perturbative interactions of the ``deconfined bag'' of quarks and
gluons with the true vacuum are parameterized by a bag constant $B$.
To make this state thermodynamically unfavorable at low temperatures,
the bag contribution to the pressure must be negative. Thus, when
computing the pressure of the QGP phase we subtract $B$ from the
right-hand-side of eq.~(\ref{p_idgas}). Accordingly, the energy density
receives a positive contribution, cf.~(\ref{e_idgas}), while $s$ and
$\rho_B$ remain unchanged. This additional ``bag term'' can also be
understood as an additional contribution $+Bg^{\mu\nu}$ due to the
non-perturbative interactions to the energy-momentum tensor of the QGP-fluid.

In the low-temperature region we assume an
ideal hadron gas that includes the well-established (strange and non-strange)
hadrons up to masses of $\sim2$~GeV. They are listed in Tab.~\ref{bartab}
and~\ref{mestab}. Although heavy states are rare in
thermodynamical equilibrium, they have a larger entropy per particle
than light states, and therefore have considerable impact on the
evolution. In particular, hadronization is significantly faster
as compared to the case where the hadron gas consists of light mesons
only (see the discussion in~\cite{BSch,JSoll,DumRi,feedback,CRS}).

The actual model used for the hadronic stage of the reaction (UrQMD,
see section~\ref{UrQMD_section}) additionally assumes a continuum of
color-singlet states called ``strings'' above the $m\simeq2$~GeV
threshold to model $2\rightarrow n$ processes and inelastic
processes at high CM-energy. For example, the annihilation of an
$\overline{p}$ on an $\Omega$ is described as excitation of two
strings with the same quantum numbers as the incoming hadrons,
respectively, which are subsequently mapped on known hadronic states
according to a fragmentation scheme. Since we shall be interested in
the dynamics of the $\Omega$-baryons emerging from the hadronization
of the QGP, it is unavoidable to treat string-formation.
The fact that string degrees of freedom are not taken into account
in the EoS~(\ref{p_idgas}) does not represent a problem in our case
because we focus on rapidly expanding systems where those degrees of
freedom can not equilibrate~\cite{belkbrand}.

The phase coexistence region 
is constructed employing Gibbs' conditions of phase equilibrium. The bag
parameter of $B=380$~MeV/fm$^3$ is chosen to yield the critical temperature
$T_C\approx160$~MeV at $\rho_B=0$.
By construction the EoS exhibits a first order phase transition, as
is also expected in QCD for the quark-hadron phase transition in the 
case of three light quark flavors~\cite{Pisarski:1984ms}.

The most striking aspect of a first-order phase transition with
respect to the dynamical evolution is that the pressure is almost
constant within the phase coexistence region (in fact, in a fluid
where all conserved currents vanish identically $p={\rm const.}$
within the mixed phase). Thus, the isentropic speed of sound,
\be
c_S^2 = \frac{\PD p}{\PD \epsilon}\Big|_{s/\rho_B}\quad,
\ee
is very small. This quantity characterizes the pressure gradient
caused by a given energy density gradient along an isentrope, i.e.\
at constant entropy density per net baryon density (recall
that all continuous solutions of the relativistic ideal-fluid dynamical
equations conserve the entropy). A very small $c_S$ means that
(isentropic) expansion is inhibited because the fluid does not
``respond'' to energy density gradients. In heavy-ion collisions
this reflects in a particularly ``soft'' expansion if the mixed phase occupies
the largest space-time volume of all three
phases~\cite{DumRi,RiGy2,kataja,RiGy96a,soft}.
For recent discussions
of the consequences of this effect in cosmology (primordial black hole
formation, evolution of density perturbations through the QCD phase
transition) see e.g.~\cite{vs_cosm}.

However, a nearly vanishing isentropic velocity of sound does only
occur if the net baryon density is not very large, as e.g.\ in the
cosmological QCD phase transition or in the central region of high-energy
collisions studied here. In heavy-ion collisions at much lower
energies, where the net baryon density in the central region is rather large,
$c_S$ is not very small. Despite the first-order phase transition, the
isentropic expansion of baryon-dense fluids is not inhibited~\cite{aflow40}.

One should also be aware of the fact that by constructing the phase coexistence
region with Gibbs' conditions we implicitly assume a ``well-mixed'' phase,
i.e.\ that the transition from the QGP to the hadronic stage proceeds
in equilibrium. This is the common approach widely employed in the
literature~\cite{BSch,JSoll,DumKat,DumRi,hung98a,feedback,bassdum99,dumitru99a,RiBe,RiGy2,CRS,kataja,RiGy96a,soft,aflow40,KPjp,Alam:1993jt},
and so far it is not in contradiction to existing data.
It is based on the picture that the first-order phase transition
proceeds via nucleation of hadronic bubbles in the expanding
QGP~\cite{bubble_nucl}, and
that the bubble nucleation and growth is fast as compared to the expansion
rate such that the two phases are approximately in pressure equilibrium.
However, this scenario is less likely to apply to high-energy heavy-ion
collisions than to the cosmological QCD phase transition, because in the
former case the
expansion rate is many order of magnitude larger~\cite{ad99}.
In particular, it has been speculated recently that the time-scale for
supercooling down to the spinodal instability is comparable to that for
homogeneous bubble nucleation~\cite{HeisJack}. Thus, it may well
be that the phase transition proceeds via spinodal decomposition rather
than bubble nucleation. In that case, the ``soft'' mixed phase with
$c_s^2\approx0$ would be
absent and shorter reaction times may be expected.
In any case, we postpone a detailed dynamical study of the latter
scenario to a future publication, and shall restrict ourselves here to the
more conservative picture assuming an adiabatic phase transition.

Finally, we have to specify the initial conditions:
\begin{itemize}
\item[\bf SPS:]
For collisions at SPS energy we assume that hydrodynamic flow sets in
on the hyperbola $\tau_i=1$~fm/c. This is a value conventionally
assumed in the literature, cf.\ e.g.~\cite{Bj}.
We further employ a (net) baryon rapidity density (at mid-rapidity) of
$dN_B/dy=80$, as obtained by the NA49-collaboration for central Pb+Pb
reactions~\cite{NA49netB}. The average specific entropy in these collisions is
$\overline{s}/\overline{\rho}_B=45\pm5$ (the bar indicates averaging over
the transverse plane).
That entropy per net baryon fits most measured hadron
multiplicity ratios within
$\pm20\%$~\cite{bms}.
The corresponding initial energy and net baryon densities
($\overline{\epsilon}_i=6.1$~GeV/fm$^3$, $\overline{\rho}_i=4.5\rho_0$)
are assumed to be distributed in the transverse
plane according to a so-called ``wounded nucleon'' distribution
with transverse radius $R_T=6$~fm, i.e.\
$\epsilon(\tau_i,r_T), \rho_B(\tau_i,r_T) \propto f(r_T)$,
with $f(r_T) = \frac{3}{2}\sqrt{1-r_T^2/R_T^2}$.
The initial temperature and quark-chemical potentials are
(they are of course not exactly constant over the transverse plane)
$T_i\simeq220$~MeV, $\mu_q\simeq150$~MeV, $\mu_s=0$.
The transverse velocity field on the $\tau=\tau_i$ hyperbola
is assumed to vanish.

\item[\bf RHIC:]
Due to the higher parton
density at mid-rapidity as compared to collisions at SPS energy,
thermalization may be reached earlier at RHIC. According to various
studies~\cite{Biro:1993qt,T0}, thermalization might occur within
$\sim0.3-1$~fm. We assume $\tau_i=R_T/10=0.6$~fm.
The net baryon rapidity density and specific entropy
at mid-rapidity in central Au+Au at $\sqrt{s}=200A$~GeV is predicted by
various models of the initial evolution, e.g.\ the parton cascade model
{\small PCM}, {\small RQMD~1.07}, {\small FRITIOF~7}, and {\small HIJING/B},
to be in the range $\d N_B/\d y\approx20-35$,
$s/\rho_B\approx150-250$~\cite{GeigKap,microRHIC}. We will employ
$\d N_B/\d y=25$ and $\overline{s}/\overline{\rho}_B=205$
($\rightarrow\overline{\epsilon}_i=20$~GeV/fm$^3$,
$\overline{\rho}_i=2.3\rho_0$).
These parameters could of course be fine-tuned once the first experimental
data are available.
As in the above case, $\epsilon(\tau_i,r_T)$ and
$\rho_B(\tau_i,r_T)$ are initially distributed in the
transverse plane according to a wounded-nucleon distribution with
$R_T=6$~fm.
The initial temperature and quark-chemical potentials
follow as $T_i\simeq300$~MeV, $\mu_q\simeq45$~MeV, $\mu_s=0$, respectively.
This corresponds to a transverse energy on the $\tau=\tau_i$ hyperbola
of $\d E_T/\d y\simeq1.3$~TeV, which decreases to
$\d E_T/\d y\simeq720$~GeV on the
hadronization hypersurface~\cite{bassdum99,dumitru99a}.

\item[\bf LHC:]
The initial conditions for CERN-LHC energy are, of course, less well
known. Qualitatively, and according to present expectations, it appears
reasonable to assume that
\begin{enumerate}
\item the density of minijets produced at time $\tau_0\sim 1/p_0$, where
$p_0\sim 2$~GeV is the minijet cut-off scale, is much larger than at
BNL-RHIC energy. The most recent estimates of the energy densities in the
central region span the range $\epsilon_0=(0.3-1.3)$~TeV/fm$^3$
\cite{Hammon:1999vw,Emel'yanov:1999bn}.
The results to be expected from $p+p$, $p+A$, and $A+A$ at BNL-RHIC will
probably not reduce the uncertainties by much because the energy density
at $y\sim0$ and $\sqrt{s}=5.5A$~TeV depends strongly on the model for the
nuclear parton distribution functions at very small $x$, out of range
for RHIC.
\item the higher initial density of partons could also lead to somewhat
faster equilibration than at the lower energies. Note that the produced
gluons already have the ``right'' thermal energy per particle,
$\epsilon_0/\rho_0\simeq 2.7 T_0$ \cite{Emel'yanov:1999bn}.
The distribution in momentum-space, however, has to become isotropic via
rescattering among the partons~\cite{Gyulassy:1997ib,GeigKap}.
\item the net baryon charge in a rapidity-slice $\Delta y=1$
around $y=0$ is even smaller than at RHIC, and can in practice be neglected
if one focuses on the bulk dynamics of the central region (in the same way
as we neglect net strangeness, charm, etc.). Note however,
that smaller rapidity bins may exhibit quite large fluctuations of the
initial net baryon charge~\cite{Spieles:1996is}.
\end{enumerate}
Thus, in view of these uncertainties, it is clear that precise
quantitative predictions for the CERN-LHC energy are hardly possible at the
moment. Our more modest aim will therefore be to discuss a set of
even more ``extreme'' initial conditions than those employed for CERN-SPS and
BNL-RHIC energies, to give an idea how the dynamical evolution may continue
at even higher energies. Whether or not that set of initial conditions
corresponds closely to the LHC case can not be decided presently
on solid grounds.

Thus, we employ a thermalization time $\tau_i=0.3$~fm, an initial
energy density (not on the $\tau_0$ but on the $\tau_i$ hypersurface~!)
$\overline{\epsilon}_i=230$~GeV/fm$^3$, and a vanishing net baryon charge,
$\d N_B/\d y=\rho_i=0$. Again, the initial energy density is distributed
in the transverse plane according to a wounded nucleon distribution with
$R_T=6$~fm. The initial temperature is about $T_i\simeq580$~MeV (it is
not exactly constant over the transverse plane), the initial
transverse energy is $\d E_T/\d y\simeq7.8$~TeV.
\end{itemize}

\subsection{Hadronization and the transition to microscopic dynamics}

Having specified the initial conditions on the $\tau=\tau_i$ hypersurface
and the
EoS, the hydrodynamical solution in the forward light-cone is determined
uniquely. As already mentioned in section~\ref{FD_MT_general}, we assume
that it is not a bad approximation to determine the hadronization
hypersurface a posteriori from the solution in the whole forward light-cone.
In other words, the hadronization hypersurface is assumed to be within the
region of validity of hydrodynamics.

In parametric representation, the hypersurface $\sigma^\mu$
is a function of three parameters~\cite{MTW_G}.
In our case, due to the symmetry under rotations around and Lorentz-boosts
along the beam axis, two of these parameters can simply be identified with
$\eta$ and $\phi$, while $\tau$ and $r_T$ depend only on the third
parameter, call it $\zeta$. Thus, $\zeta\in\left[0,1\right]$ parameterizes the
hypersurface in the planes of fixed $\eta$ and $\phi$ (in the mathematically
positive orientation, i.e.\ counter clock-wise).
The normal is~\cite{MTW_G}
\bea
\label{dsigmu}
\d\sigma_\mu &=& \epsilon_{\mu\alpha\beta\gamma}
      \frac{\PD\sigma^\alpha}{\PD \zeta}
      \frac{\PD\sigma^\beta}{\PD \eta}
      \frac{\PD\sigma^\gamma}{\PD \phi}
\d\zeta\d\eta\d\phi \nonumber\\
&=& \left( -\frac{{\d}r_T}{{\d}\zeta}\, \cosh
\eta, \, \frac{{\d}\tau}{{\d}\zeta}\, \cos \phi,\, 
\frac{{\d}\tau}{
{\d}\zeta}\, \sin \phi, \, \frac{{\d}r_T}{{\d}\zeta}\, \sinh \eta
\right)\, r_T\, \tau \, \, {\d} \zeta\, {\d} \eta\,
{\d} \phi\,\, .
\eea
This expression naturally looks simpler in the
$(\tau,\eta,r_T,\phi)$-basis, cf.\ e.g.~\cite{ad99}, but we will
nevertheless write all vectors and tensors in the $(t,x,y,z)$-basis
throughout the manuscript, even if the components are written in terms
of the variables $\tau$, $\eta$, $r_T$, and $\phi$.

We can now apply eq.~(\ref{CF_gen}) to compute
the number of hadrons of species $i$ hadronizing at space-time rapidity
$\eta$, proper time $\tau$, and
position $r_T\left(\cos\left(\chi-\phi\right),\sin\left(\chi-\phi\right)
\right)$,
with four-momentum $p^\mu=(m_T\cosh y,p_T\cos\chi,
p_T\sin\chi,m_T\sinh y)$, 
\be \label{hadr_distr}
\frac{\d^6N_i}{d^2p_T\d y\d\eta \d\zeta \d\phi} =
r_T\,\tau\left( p_T\cos\left(\chi-\phi\right)
\frac{d\tau}{d\zeta}
- m_T\cosh(y-\eta)\frac{dr_T}{d\zeta}\right)\,
f_i\left(p\cdot u\right) \quad.\label{CFspectra}
\ee
$u^\mu=\gamma_T (\cosh\eta,v_T\cos(\chi-\phi),v_T\sin(\chi-\phi),
\sinh\eta)$ denotes the fluid four-velocity.
Thus, the direction of the particle momentum in the
transverse plane is determined by the angle $\chi$, while the relative
angle between $\vec{p}_T$ and the transverse flow velocity, $\vec{v}_T$,
is denoted by $\phi$.
$f$ is either a Bose-Einstein or Fermi-Dirac distribution function, depending
on the particle species under consideration.

How is the distribution~(\ref{hadr_distr}) actually passed to the
microscopic model~? First, it is integrated over space-time
($\eta$, $\zeta$, $\phi$) and momentum space ($\vec{p}_T$, $y$),
rounded to an integer value (the hadronic transport model described
in the next section deals with integer number of particles, only),
and the distribution~(\ref{hadr_distr}) divided by $N_i$ is used as
probability distribution to randomly generate space-time and
momentum-space coordinates for $N_i$ hadrons of species $i$.
Of course, due to the fact that our system has a surface and does not
extend to infinity in the transverse plane, hadronization does not occur on
a $\tau={\rm const.}$ hypersurface, cf.\ Fig.~\ref{h_hyper}. Thus,
if we look at our expanding system on  $\tau={\rm const.}$ surfaces,
there exists an interval where the two models, hydrodynamics and
the microscopic transport, are applied in parallel.

\subsection{Microscopic dynamics: the UrQMD approach}
\label{UrQMD_section}
The ensemble of hadrons generated accordingly is
then used as initial condition for the microscopic transport model
Ultra-relativistic Quantum Molecular Dynamics (UrQMD) 
\cite{uqmdref1}. The UrQMD approach is closely related to hadronic 
cascade \cite{cascade}, Vlasov--Uehling--Uhlenbeck \cite{vuu}
and (R)QMD transport models \cite{qmd}.
We shall describe here only the part of the model that is important
for the application at hand, namely the evolution of an expanding hadron gas
in local equilibrium at a temperature of about $T_C\sim160$~MeV.
The treatment of high-energy hadron-hadron scatterings, as it occurs in
the initial stage of ultrarelativistic collisions, is not discussed here.
A complete description of the model and detailed comparisons to experimental
data can be found in~\cite{uqmdref1}.

The basic degrees of freedom are hadrons modeled as Gaussian wave-packets,
and strings, which are used to model the fragmentation of high-mass
hadronic states via the Lund scheme \cite{lund}.
The system evolves as a sequence of binary collisions or $2-N$-body decays 
of mesons, baryons, and strings.

The real part of the nucleon optical potential, i.e.\ a mean-field,
can in principle be included in UrQMD for the dynamics of baryons 
(using a Skyrme-type interaction with a hard equation of state). However,
currently no mean field for mesons (the most abundant hadrons in our
investigation) are implemented.
Therefore, we have not accounted for mean-fields in the equation
of motion of the hadrons. 
To remain consistent, mean fields were also not taken into account
in the EoS on the fluid-dynamical side. Otherwise, pressure equality (at given
energy and baryon density) would be destroyed.
We do not expect large modifications of the results presented here due to
the effects of mean fields, since the ``fluid'' is not very dense after 
hadronization and current experiments at SIS and AGS only point
to strong medium-dependent properties of mesons (kaons in particular)
for relatively low incident beam energies 
($E_{lab} \le 4$~GeV/nucleon) \cite{gsikaonen}. 
Nevertheless, mean fields will have to be included in the future;
a fully covariant treatment of baryon and meson dynamics within UrQMD
derived from a chiral Lagrangian~\cite{zschiesche} is currently
under development.
 
Binary collisions are performed in a point-particle sense:
Two particles collide if their minimum distance $d$, 
i.e.\ the minimum relative 
distance of the centroids of the Gaussians during their motion, 
in their CM frame fulfills the requirement: 
\begin{equation}
 d \le d_0 = \sqrt{ \frac { \sigma_{\rm tot} } {\pi}  }  , \qquad
 \sigma_{\rm tot} = \sigma(\sqrt{s},\hbox{ type} ).
\end{equation}
The cross section is assumed to be the free cross section of the
regarded collision type ($N-N$, $N-\Delta$, $\pi-N$ \ldots).

The UrQMD collision term contains 53 different baryon species
(including nucleon, delta and hyperon resonances with masses up to 2 GeV) 
and 24 different meson species (including strange meson resonances), which
are supplemented by their corresponding anti-particle 
and all isospin-projected states.
The baryons and baryon-resonances which can be populated in UrQMD are listed
in table~\ref{bartab}, the respective mesons in table~\ref{mestab} -- 
full baryon/antibaryon symmetry is included (not shown in the table), both,
with respect to the included hadronic states, as well as with respect to
the reaction cross sections.
All hadronic states can be produced in string decays, s-channel
collisions or resonance decays. 

Tabulated and parameterized experimental 
cross sections are used when available. Resonance absorption, decays 
and scattering are handled via the principle of detailed balance. 
If no experimental information is
available, the cross section is either  calculated via
an One-Boson-Exchange (OBE) model or via a modified additive quark model
which takes basic phase space properties into account.

In the baryon-baryon sector, the total and elastic proton-proton and 
proton-neutron cross sections are well known \cite{PDG96}. 
Since their functional dependence on $\sqrt{s}$ shows
a complicated shape at low energies, UrQMD uses a table-lookup for those
cross sections. However, many cross sections involving strange baryons and/or 
resonances are not well known or even experimentally accessible -- for these
cross sections the additive quark model is widely used.

As we shall see later, the most important reaction channels 
in our investigation are meson-meson and
meson-baryon elastic scattering and resonance formation. 
For example, the total meson-baryon cross section for
non-strange particles is  given by
\begin{eqnarray}
\label{mbbreitwig}
\sigma^{MB}_{tot}(\sqrt{s}) &=& \sum\limits_{R=\Delta,N^*}
       \langle j_B, m_B, j_{M}, m_{M} \| J_R, M_R \rangle \,
        \frac{2 S_R +1}{(2 S_B +1) (2 S_{M} +1 )} \nonumber \\
&&\times        \frac{\pi}{p^2_{CMS}}\, 
        \frac{\Gamma_{R \rightarrow MB} \Gamma_{tot}}
             {(M_R - \sqrt{s})^2 + \frac{\Gamma_{tot}^2}{4}}
\end{eqnarray}
with the total and partial $\sqrt{s}$-dependent decay widths $\Gamma_{tot}$ and
$\Gamma_{R \rightarrow MB}$. 
The full decay width $\Gamma_{tot}(M)$ of a resonance is 
defined as the sum of all partial decay widths and depends on the
mass of the excited resonance:
\begin{equation}
\Gamma_{tot}(M)  
\label{gammatot}
       \,=\, \sum  \limits_{br= \{i,j\}}^{N_{br}} \Gamma_{i,j}(M) \quad.
\end{equation}
The partial decay widths $\Gamma_{i,j}(M)$ for the decay into the 
final state with particles $i$ and $j$ is given by
\begin{equation}
\label{gammapart}
\Gamma_{i,j}(M)
        \,=\,
       \Gamma^{i,j}_{R} \frac{M_{R}}{M}
        \left( \frac{\langle p_{i,j}(M) \rangle}
                    {\langle p_{i,j}(M_{R}) \rangle} \right)^{2l+1}
         \frac{1.2}{1+ 0.2 
        \left( \frac{\langle p_{i,j}(M) \rangle}
                    {\langle p_{i,j}(M_{R}) \rangle} \right)^{2l} } \quad,
\end{equation}
here $M_R$ denotes the pole mass of the resonance, $\Gamma^{i,j}_{R}$
its partial decay width into the channel $i$ and $j$ at the pole and
$l$ the decay angular momentum of the final state.
All pole masses and partial decay widths at the pole are taken from the Review
of Particle Properties \cite{PDG96}. 
$\Gamma_{i,j}(M)$ is constructed in such a way that 
$\Gamma_{i,j}(M_R)=\Gamma^{i,j}_R$ is fulfilled at the pole.
In many cases only crude estimates for $\Gamma^{i,j}_R$ are given
in \cite{PDG96} -- the partial decay widths must then be fixed by
studying exclusive particle production in elementary proton-proton
and pion-proton reactions. 
Therefore, e.g., the total pion-nucleon cross section depends on the
pole masses, widths and branching ratios of all $N^*$ and $\Delta^*$
resonances listed in table~\ref{bartab}. 
Resonant meson-meson scattering
(e.g. $\pi + \pi \to \rho$ or $\pi + K \to K^*$)
is treated in the same formalism.

In order to correctly treat equilibrated matter~\cite{belkbrand}
(we repeat that the hadronic
matter with which UrQMD is being initialized in our approach is in local
chemical and thermal equilibrium), the principle of detailed balance is
of great importance.
Detailed balance is based on time-reversal invariance 
 of the matrix element of the reaction. It is most commonly found
in textbooks in the form:
\begin{equation}
\label{dbgl3}
\sigma_{f \rightarrow i } \,=\, \frac{\vec{p}_i^2}{\vec{p}_f^2} \frac{g_i}{g_f}
\sigma_{i \rightarrow f} \quad ,
\end{equation}
with $g$ denoting the spin-isospin degeneracy factors.
UrQMD applies the general principle of detailed balance to the 
following two process classes:
\begin{enumerate}
\item 
Resonant meson-meson and meson-baryon interactions: Each resonance created
via a meson-baryon or a meson-meson annihilation may again decay into
the two hadron species which originally formed it. This symmetry is only
violated in the case of three- or four-body decays and string fragmentations, 
since N-body collisions with (N$>2$) are not implemented in UrQMD. 
\item
Resonance-nucleon or resonance-resonance interactions: the excitation
of baryon-resonances in UrQMD is handled via parameterized cross sections
which have been fitted to data. The reverse reactions usually have not
been measured - here the principle of detailed balance is applied.
Inelastic baryon-resonance de-excitation is the only method in UrQMD
to absorb mesons (which are {\em bound} in the resonance). Therefore
the application of the detailed balance principle is of crucial
importance for heavy nucleus-nucleus collisions.
\end{enumerate}

Equation~(\ref{dbgl3}), however, is only valid in the case of stable
particles with well-defined masses. Since in UrQMD detailed balance
is applied to reactions involving resonances with finite lifetimes
and broad mass distributions, equation~(\ref{dbgl3}) has to be 
modified accordingly. For the case of one incoming resonance the
respective modified detailed balance relation has been derived
in \cite{danielewicz91a}. Here, we generalize this expression for
up to two resonances in both, the incoming and the outgoing channels.

The differential cross section for the reaction 
$(1\,,\,2) \rightarrow (3\,,\,4)$ is given by:
\begin{equation}
\label{diffcx1}
{\rm d} \sigma_{12}^{34} \,=\,
    \frac{| {\cal M} |^2}{64 \pi^2 s} \, \frac{p_{34}}{p_{12}} 
        \,{\rm d \Omega}\,
     \prod_{i=3}^4    \delta(p_i^2 -M_i^2) {\rm d}p_i^2  \quad,
\end{equation}
here the $p_i$ in the $\delta$-function denote four-momenta.
The $\delta$-function ensures that the particles are on mass-shell,
i.e. their masses are well-defined. If the particle, however, has 
a broad mass distribution, then the $\delta$-function
must be substituted by the respective mass distribution (including
an integration over the mass):
\begin{equation}
\label{diffcx2}
{\rm d} \sigma_{12}^{34} \,=\,
    \frac{| {\cal M} |^2}{64 \pi^2 s} \, \frac{1}{p_{12}} 
        \,{\rm d \Omega}\,
     \prod_{i=3}^4  p_{34} \cdot
   \frac{ \Gamma}{\left(m-M_i\right)^2+ \Gamma^2/4} \frac{{\rm d} m}{2\pi}
\, .
\end{equation}
Incorporating these modifications into equation~(\ref{dbgl3}) and
neglecting a possible mass-dependence of the matrix element we
obtain:
\begin{equation}
\label{uqmddetbal}
     \frac{ {\rm d} \sigma_{34}^{12} }{{\rm d} \Omega }   
       = \frac{\langle p_{12 }^2 \rangle   }
       {\langle p_{34 }^2 \rangle  } \,
       \frac{(2 S_1 + 1) (2 S_1 + 1)}
            {(2 S_3 + 1) (2 S_4 + 1)}\,
      \sum_{J=J_-}^{J_+} 
      \langle j_1 m_1 j_2 m_2 \| J M \rangle \, 
        \frac{ {\rm d} \sigma_{12}^{34} }{{\rm d} \Omega } \, .
\end{equation}
Here, $S_i$ indicates the spin of particle $i$ and 
the summation of the Clebsch-Gordan-coefficients is over the isospin of the
outgoing channel only. For the incoming channel, isospin is 
treated explicitly. The summation limits are given by:
\begin{eqnarray}
  J_- &=& \max \left( |j_1-j_2|,  |j_3-j_4| \right) \\  
  J_+ &=& \min \left( j_1+j_2,  j_3+j_4     \right)  \quad.
\end{eqnarray}
The integration over the mass distributions of the resonances  
in equation~(\ref{uqmddetbal}) has been denoted by the brackets 
$\langle\rangle$,
e.g.
\begin{displaymath}
p_{3,4}^2 \,\Rightarrow \, \langle p_{3,4}^2 \rangle \, = \,
 \int  \int 
  p_{CMS}^2(\sqrt{s},m_3,m_4)\, A_3(m_3) \, A_4(m_4) \, 
  {\rm d} m_3\; {\rm d} m_4 \quad,
\end{displaymath}
with the  
mass distribution $A_r(m)$ given by a free Breit-Wigner distribution
with a mass-dependent width according to equation~(\ref{gammatot}):
\begin{equation}
\label{bwnorm}
A_r(m) \, = \, \frac{1}{N} 
        \frac{\Gamma(m)}{(m_r - m)^2 + \Gamma(m)^2/4} \qquad \mbox{with} \quad
        \lim_{\Gamma \rightarrow 0} A_r(m) = \delta(m_r - m) \, ,
\end{equation}
with the normalization constant
\begin{equation}
\label{bwnorm_norm}
N \,=\, \int\limits_{-\infty}^{\infty} 
\frac{\Gamma(m)}{(m_r - m)^2 + \Gamma(m)^2/4} \, {\rm d}m \, .
\end{equation}
Alternatively one can also choose a Breit-Wigner distribution with a fixed
width, the normalization constant then has the value $N=2\pi$.

The most frequent applications of equation~(\ref{uqmddetbal}) in UrQMD
are the processes $\Delta_{1232}\, N \to N\, N$ and
$\Delta_{1232}\, \Delta_{1232} \to N\, N$.

\section{Results for heavy-ion collisions at CERN-SPS, BNL-RHIC and CERN-LHC}
We now present some representative results for central collisions of heavy
ions at CERN-SPS, BNL-RHIC and CERN-LHC energies. We will focus on
single-inclusive momentum-space distributions and the space-time
picture of freeze-out following the hadronization phase transition.
As we shall see, one already gains much insight into the dynamics of
high-energy heavy-ion collisions from these observables. Many other
aspects are thinkable and interesting but have to be postponed to future
studies.

\subsection{Hydrodynamical expansion and hadronization}

We first briefly discuss the evolution and hadronization of the
QGP-cylinder present on the $\tau=\tau_i$ hypersurface as obtained
from the hydrodynamical solution. Similar arguments and results
can be found in a variety of papers, see for
example~\cite{BSch,JSoll,hung98a,RiBe,CRS,kataja,RiGy96a,Alam:1993jt}.

In particular, ref.~\cite{DumRi} employed the very same model as here
(i.e.\ longitudinal scaling
flow with cylindrically symmetric transverse expansion, the
initial conditions and the EoS).
However, the evolution at CERN-LHC energy had not
been covered, and the hadronization hypersurface was only shown for
a step-function like initial transverse energy density distribution,
but not for the wounded-nucleon distribution employed here.
Therefore, a short discussion of the prehadronic stage may be in order here.

\begin{figure}
\centerline{\epsfig{figure=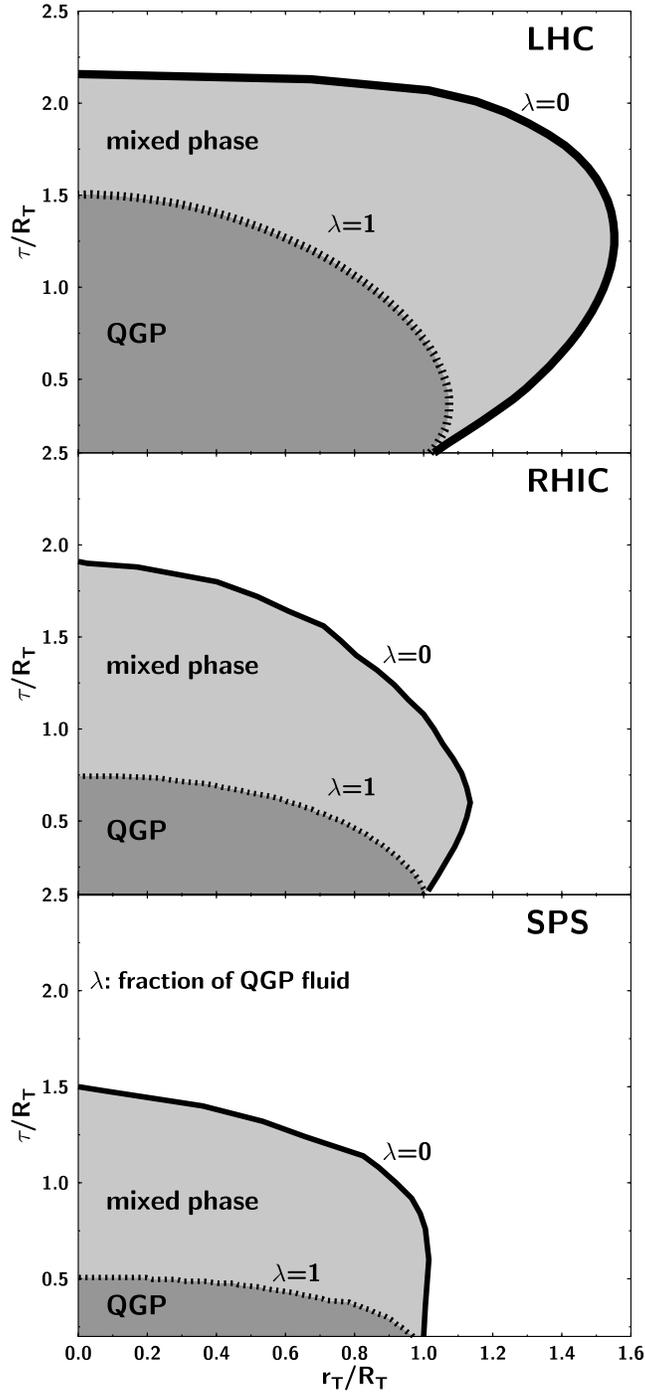,width=3.5in}}
\caption{\label{hfs} Hypersurfaces corresponding to $\lambda=1$ 
(boundary between pure QGP and mixed phase)
and $\lambda=0$ (boundary between mixed phase and pure hadron phase)
for LHC (top), RHIC (middle) and SPS (bottom). }
\label{h_hyper}
\end{figure}

Fig.~\ref{h_hyper} summarizes the space-time picture in the plane
$\eta=\phi=0$. We show projections of various hypersurfaces on the
$(\tau,r_T)$-plane because their shape in the $\phi$- and $\eta$-directions
is trivial: they are simply horizontal lines in the
$(\tau,\phi)$- $(\tau,\eta)$-planes, extending from $-\pi$ to $\pi$ and
$-\infty$ to $\infty$, respectively. Thus, no derivatives like
$\PD\tau/\PD\phi$ etc.\ appear in $\d\sigma_\mu$, eq.~(\ref{dsigmu}).

Basically, we start at $\tau=\tau_i$
with a pure QGP extending from
$r_T=0$ up to $r_T=R_T\equiv 6$~fm in the transverse direction.
The thickness of the non-QGP region at the surface is
very small for the wounded nucleon distribution.
Initially, the hot quark-gluon fluid is cooled mainly due
to the longitudinal expansion, except close to the surface,
where transverse pressure gradients are also large and lead
to expansion rates several times larger than the simple
$1/\tau$ law~\cite{hung98a,ad99}. The fluid eventually reaches
the boundary to the mixed phase, denoted by $\lambda=1$.
($\lambda$ is the local fraction of quarks and gluons
within the mixed phase.) Clearly, the space-time volume
of pure QGP increases substantially from SPS to RHIC and then
again towards LHC. This leads to stronger
transverse flow of matter entering the mixed phase at RHIC and LHC
than at SPS. Due to this effect the hadronization hypersurface ($\lambda=0$)
extends to larger $r_T$. 

At SPS, the hadronization hypersurface $\lambda=0$,
where the switch to the microscopic model is performed,
is almost stationary for some time $\tau\simeq R_T$, after which
the entire fluid hadronizes rapidly. UrQMD is being fed with
hadrons from the stationary surface of a ``burning log''
of mixed phase matter~\cite{RiGy96a}. This is not to be
misunderstood as an evaporation process, though. The fluid is
moving with substantial velocity through the hadronization
hypersurface, in particular near the point where the $\lambda=0$ and
the $\tau=\tau_i$ hypersurfaces meet (there, the very dilute fluid comes
close to the light-cone). Thus, the momenta of the
emitted hadrons, which are purely thermal in the local rest frame,
are boosted in transverse direction.
At RHIC, and of course even more so at LHC, that pre-acceleration
by the QGP ``explosion'' is so strong that the hadronization
hypersurface is initially even driven outwards, before the
mixed-phase cylinder finally collapses (when it
can not balance the vacuum pressure $B$ any more)
and emits hadrons from all over the transverse area. 

Thus, it is clear from Fig.~\ref{h_hyper} that the dynamics at
SPS is characterized by the large space-time volume occupied by
the mixed phase, while the stiffer and more ``explosive''
\cite{RiGy96a,SZ} QGP gains importance at higher energies.

\subsection{Post-hadronization kinetics: evolution of $\langle p_T\rangle$}
\label{meanptevol}
The choice of the hypersurface at which to perform the
transition from the macroscopic hydrodynamical
calculation to the microscopic transport model may affect the reaction
dynamics and the results of our calculation.
However, concerning the variation of the hypersurface for that transition 
one has to note that the hadronic 
part of the EoS used in the hydrodynamic solution contains the same states as 
UrQMD, and the energy-momentum tensors on both sides of the hadronization 
hypersurface match. 
{\em If} the assumption of local equilibrium is indeed fulfilled, 
UrQMD will simply continue the hydrodynamic flow since
it reduces to hydrodynamics in the equilibrium-limit. 
However, as we shall see later, 
for some hadron species with small interaction cross sections 
deviations from ideal hydrodynamic flow can be observed immediately after 
complete hadronization (see also  refs. \cite{bassdum99,dumitru99a}).
It is found that the expansion of the 
hadronic fluid is dissipative rather than ideal; due to the fast local 
expansion  generated by the QGP before hadronization the ideal flow is 
disturbed. Therefore it does not make much sense to choose a later
hypersurface for the matching because one would precisely assume that ideal 
flow persists even after hadronization.

\begin{figure}
\centerline{\epsfig{figure=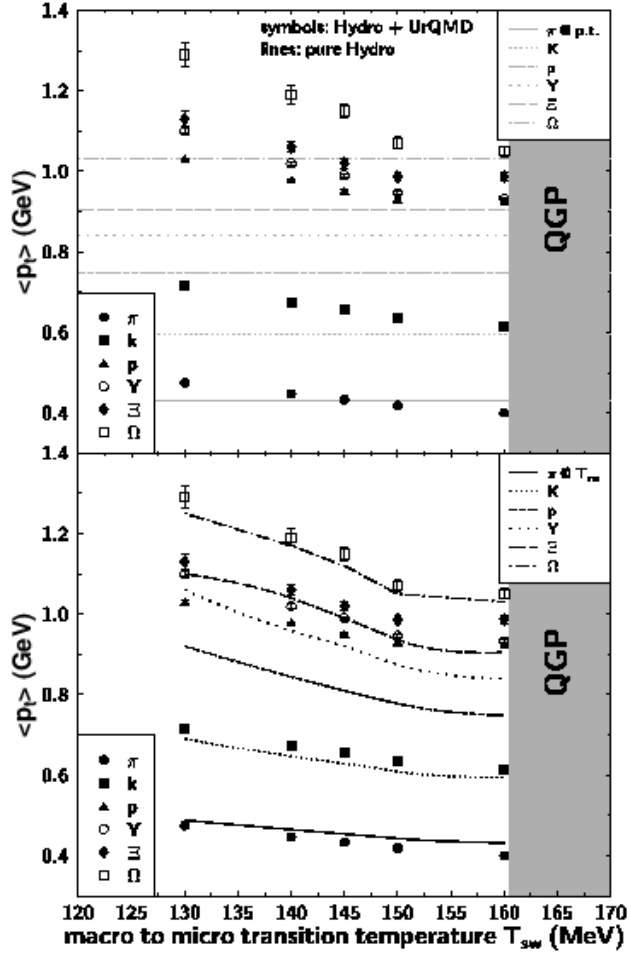,width=3.5in}}
\caption{\label{ptt3} Mean transverse momentum $\langle p_T \rangle$ 
of various hadron species at freeze-out (symbols) vs.\ the macro to micro
transition temperature $T_{sw}$. The horizontal lines in the upper frame show
the respective $\langle p_T \rangle$ values right after hadronization
($\lambda=0$ hypersurface). The lines in the lower frame show the
$\langle p_T \rangle$ emerging from ideal flow down to $T=T_{sw}$.
This figure is for central $Au+Au$ collisions at BNL-RHIC energy.}
\end{figure}

Nevertheless, it is interesting to study how the choice of a later
hypersurface for the transition from the macroscopical to the microscopical
part of the calculation affects the results. This reveals ``how wrong''
the assumption of an ideal evolution of the state at hadronization is.
Figure~\ref{ptt3} shows the final mean transverse momentum $\langle p_T\rangle$
for various hadron species as a function of the temperature on the 
hydro$\rightarrow$micro transition isotherm, $T_{sw}$.
The grey lines in 
the upper frame denote the $\langle p_T\rangle$ of the hadrons at
hadronization, i.e. at $T_{C}=160$~MeV. As shall be discussed in greater
detail in section~\ref{flowsection}, the change in $\langle p_t\rangle$
in the hadronic phase (for our ``default'' choice
$T_{sw}=T_{C}=160$~MeV) depends strongly on the individual
hadron species. Protons and hyperons gain most, the $\Omega^-$ does not
acquire any additional $\langle p_T\rangle$ at all, and pions even loose
some $\langle p_T\rangle$ due to rescattering and additional 
soft pion production.

The results change only marginally when decreasing
$T_{sw}$ to 150~MeV. Simply speaking, UrQMD reproduces the fluid-dynamical
solution down to about $T\approx150$~MeV, for central Au+Au at RHIC energy.
At this stage, fluid-dynamics predicts that the transverse rarefaction in
the hadron fluid reaches the center. Consequently, the expansion becomes
rather spherical and transverse flow increases strongly in this
``hadronic explosion''. The lower frame in Fig.~\ref{ptt3} shows the kink
in $\langle p_T\rangle(T_{sw})$ of heavy hadrons at $T_{sw}\simeq150$~MeV
predicted by ideal hydrodynamics.

Remarkably, however, the system apparently is already in a state of too
rapid expansion for this ``hadronic explosion'' to happen. Given the state
at hadronization, UrQMD (applying realistic cross-sections) predicts that the
hadronic fluid basically freezes out right at the point where the
hadronic rarefaction is about to make the expansion more spherical and
to increase the expansion rate, see e.g.\ Fig.~2 in~\cite{ad99}.
Any later transition from hydrodynamics to the microscopic transport model
leads to a strong increase of $\langle p_T\rangle$ at freeze-out,
which depends only on the mass of the
hadron, but not on its flavor (resp.\ its quark content).

The lines in the lower frame of figure~\ref{ptt3} show the $\langle p_T\rangle$
of the respective hadron species at the transition hypersurface 
(i.e. at $T_{sw}$). By comparing the $\langle p_T\rangle$ value indicated
by the line to that given by the plot symbol for each $T_{sw}$ one can
determine the amount of $\langle p_T\rangle$ gained or lost during the
microscopic evolution of the reaction. Again, protons acquire the most
$\langle p_T\rangle$ during the microscopic evolution (even though 
the amount of $\langle p_T\rangle$ gained decreases the lower $T_{sw}$ is 
and the closer the system comes to freeze-out), whereas $\Xi$'s and 
$\Omega$'s do not experience any $\langle p_T\rangle$ increase at all.

It is obvious from this analysis that the conditions of applicability for
hydrodynamics in the hadronic phase deteriorate rapidly.
A general freeze-out criterion can not be given since the freeze-out
depends on the system size and the centrality, the energy etc.
However, our transport calculation with realistic cross-sections in the
hadron gas,
starting in the wake of a hadronizing QGP, shows that the expansion
is too rapid to allow cooling of the strong interactions much below
$T_C$. In particular, adiabatic expansion breaks down once the expansion
of the hadron fluid effectively becomes (3+1)-dimensional.

\subsection{Space-time distributions of hadronic freeze-out}
\label{FO_hupersurf}
\begin{figure}
\centerline{\epsfig{figure=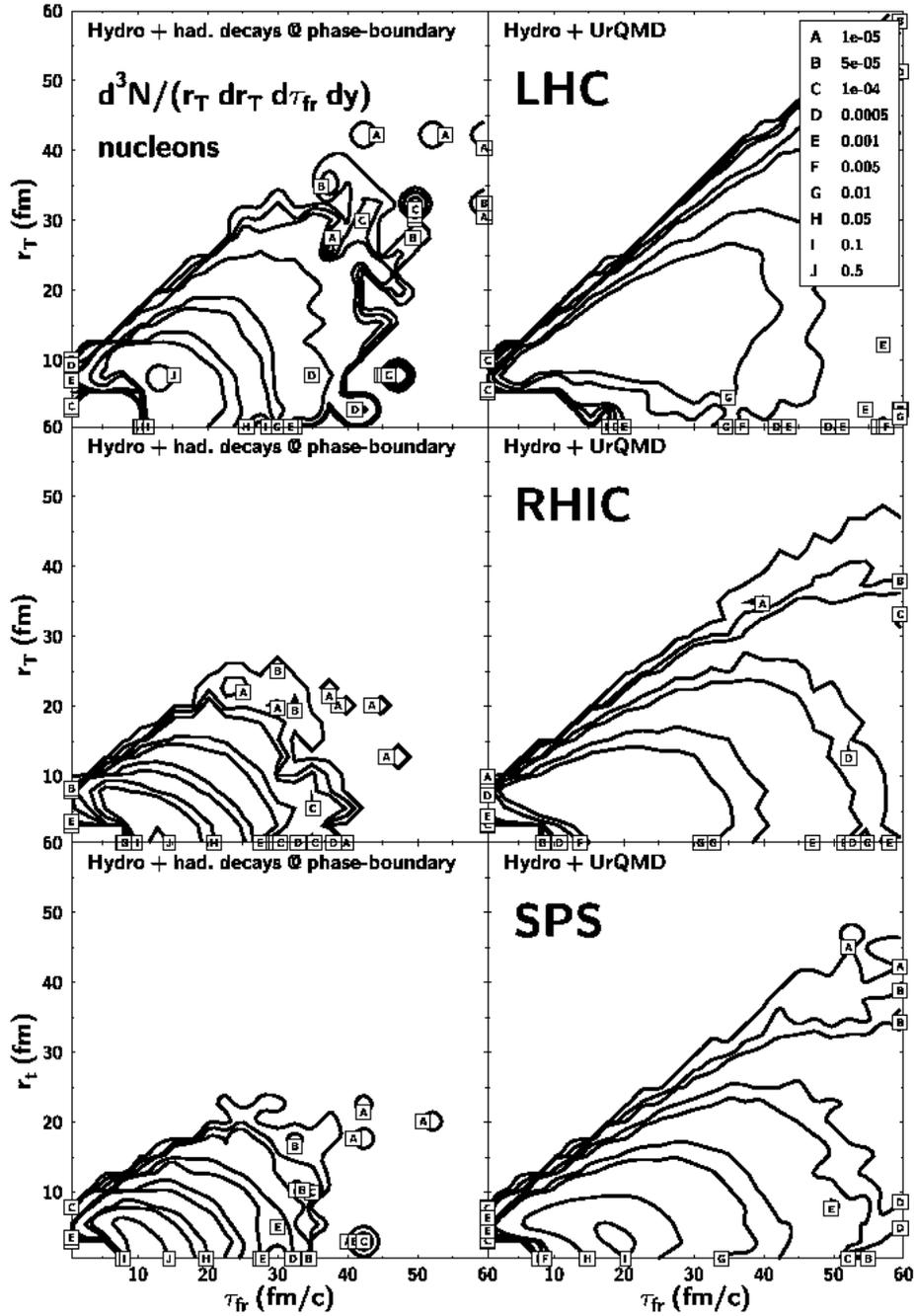,width=5in}}
\caption{\label{focontour} Freeze-out time and transverse radius distribution
d$^3N/(r_T$d$r_T$d$\tau_{fr}$d$y)$ for nucleons at LHC (top), RHIC (middle)
and SPS (bottom).
The left column shows the result for the pure hydrodynamical calculation 
up to hadronization with subsequent hadron resonance decays 
(but without hadronic reinteraction).
The right column shows the analogous calculation, but 
with full microscopic hadronic collision
dynamics after the hadronization. The contour lines
have identical binning for all rows and columns.}
\end{figure}

Let us now turn to the freeze-out ``hypersurfaces'' of pions and nucleons
in central (impact parameter 
$b = 0 $~fm) collisions of gold or lead nuclei at SPS ($\sqrt{s}=17$~GeV 
per incident colliding nucleon-pair), RHIC ($\sqrt{s}=200$~GeV 
per incident colliding nucleon-pair) and LHC ($\sqrt{s}=5500$~GeV 
per incident colliding nucleon-pair). 
We start with the nucleons, the most abundant baryon species in the system,
restricting ourselves to the
central rapidity region. Figure~\ref{focontour} shows the
freeze-out\footnote{Freeze-out meaning the space-time point of {\em last}
interaction, irrespective of how ``soft'' that last interaction might be.
We remind you also that mean-fields are not taken into account. They
could even prolong the freeze-out due to very soft interactions of the
hadrons with the mean field.}
time and transverse radius distributions 
$1/r_T$ d$^3N/$d$r_T$d$\tau_{fr}$d$y$
for LHC (top), RHIC (middle) and SPS (bottom).
The right column shows the result of the pure hydrodynamical calculation 
up to complete hadronization,
with subsequent hadronic resonance decays,
but without hadronic reinteraction. 
The left column shows the same calculation including full microscopic
hadronic dynamics.
 
The freeze-out characteristics of the nucleons are 
significantly modified due to the hadronic interaction phase. The average
transverse freeze-out radius doubles at SPS and RHIC and increases by a 
factor of 2.5 at LHC (see also table~\ref{rad_tab}).
The respective average freeze-out times increase by similar factors
(see table~\ref{tau_tab}). E.g., at RHIC the 
average freeze-out time for protons changes from
11.3 to 25.8 fm/c due to hadronic rescattering. 

As the meson multiplicity in the system at RHIC is fifty times
larger than the baryon multiplicity, baryons propagate through a
relativistic meson gas, acting as probes of this highly 
excited meson medium.
Thus, we use the proton and hyperon freeze-out values listed in 
table~\ref{tau_tab} for a first rough estimate of the duration of the 
hadronic phase via 
$\Delta \tau_{had} = \langle \tau_{fr}^{\rm Hydro+UrQMD}\rangle - 
\langle \tau_{fr}^{\rm Hydro + had. decays} \rangle $.
At the SPS $\Delta \tau_{had}$ is found to be $\approx 13.5$~fm/c,
very similar to the value at RHIC ($\approx 15$~fm/c) and at the LHC we obtain
$\Delta \tau_{had} \approx 23$~fm/c.
The transverse spatial extent of the hadronic phase can be 
estimated in a similar way, using table~\ref{rad_tab} and defining
the {\em thickness} $\Delta r_{had}$  of the hadronic phase as:
$\Delta r_{had} = \langle r_{t,fr}^{\rm Hydro+UrQMD}\rangle -
\langle r_{t,fr}^{\rm Hydro + had. decays} \rangle$.
Here we find values of $\approx 4.4$~fm at the SPS, $\approx 5.8$~fm
at RHIC and $\approx 13.3$~fm at the LHC.

The Hydro+UrQMD model predicts a space-time 
freeze-out picture which is very different from that usually 
employed in the hydrodynamical model, e.g.\ in
refs.~\cite{DumRi,locexp,hung98a,CRS,AltFO}:
Here~\cite{bassdum99}, freeze-out is found to occur
in a {\em four-dimensional} region within the forward
light-cone~\cite{Grassi97}
rather than on a three-dimensional ``hypersurface'' \cite{CF}.
Similar results have
also been obtained within other microscopic transport models~\cite{microFO}
when the initial state was not a quark-gluon plasma.
This finding seems to be a generic feature of
such models:  the elementary binary hadron-hadron interactions smear
out the sharp signals to be expected from simple hydro.
This predicted additional fourth dimension
of the freeze-out domain could affect the HBT parameters considerably. 

This does not mean that the {\em momentum-distributions} alone
can not be calculated assuming freeze-out on some effective three-dimensional
hypersurface. For example, if interactions on the outer side of that
hypersurface are very ``soft'', the single-particle momentum distributions
at not too small $p_T$ will not change anymore.
The two-particle correlator {\em does} change, however, since it probes
rather small relative momenta. Thus, the freeze-out
condition, e.g.\ the temperature, as measured by single-particle spectra
and two-particle correlations~\cite{TPFO} needs not be the same.

The shapes of the freeze-out hypersurfaces (FOHS)
show broad radial maxima for intermediate 
freeze-out times. Thus, transverse expansion has not developed scaling-flow
(in that case the FOHS would be hyperbolas in the $\tau-r_T$ plane).
This agrees with the discussion of the evolution of the $\langle p_T\rangle$
after hadronization in section~\ref{meanptevol}, which already indicated
the transition to free-streaming once the transverse expansion rate becomes
comparable to the longitudinal expansion rate.

Furthermore, the hypersurfaces of pions and nucleons, and
their shapes, are distinct from each other
(as also found in \cite{hung98a,microFO,kinDec,uqmdref1}
at the lower BNL-AGS and CERN-SPS energies). Thus,
the ansatz of a unique freeze-out hypersurface
for all hadrons appears to be a very rough approximation,
cf.\ also refs.~\cite{microFO,bassdum99,dumitru99a}.

\begin{figure}
\centerline{\epsfig{figure=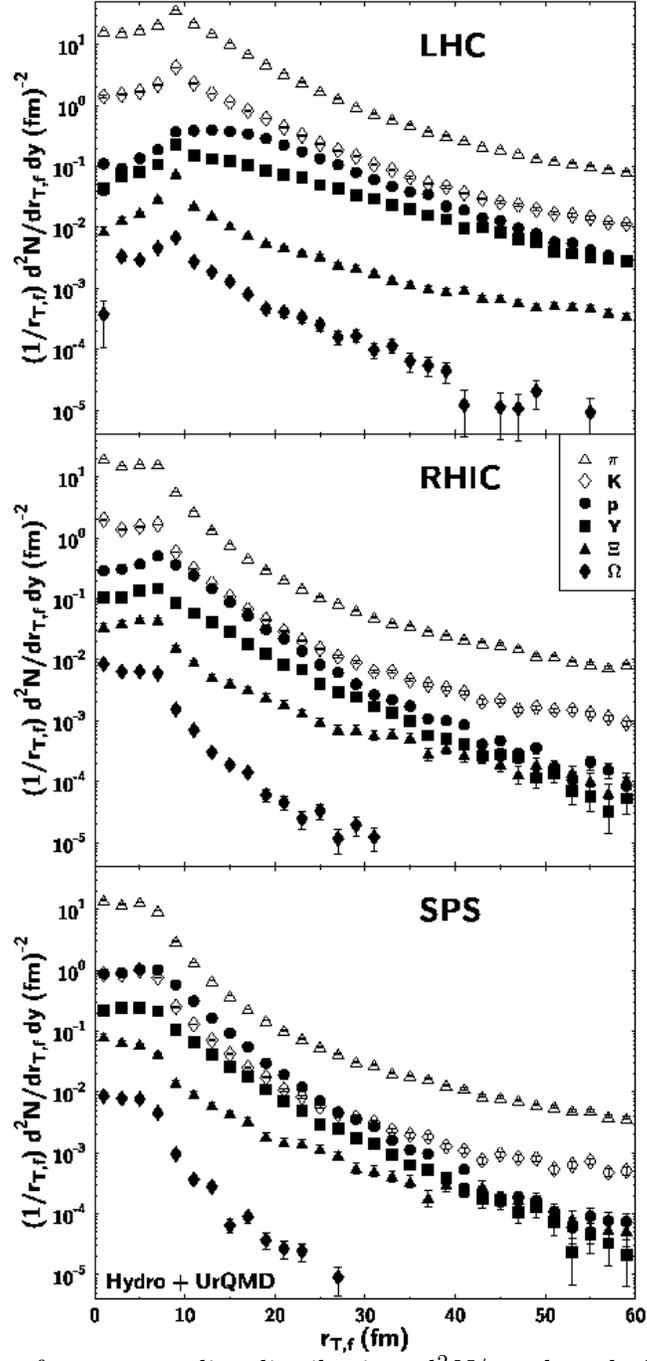,width=3.5in}}
\caption{\label{dndrt} Transverse freeze-out radius distributions
d$^2N/r_{T,f}$d$r_{T,f}$d$y$
for various hadron species at LHC (top), RHIC (middle) and SPS (bottom).}
\end{figure}

Figure~\ref{dndrt} shows the transverse freeze-out radius distributions
for $\pi$, $K$, $p$, $\Lambda+\Sigma^0$, $\Xi$ and $\Omega^-$ at LHC (top),
RHIC (middle) and SPS (bottom). 
They are rather broad and similar to each other, 
though the $\Omega^-$ shows a somewhat narrower freeze-out distribution.
The average transverse freeze-out radii are listed in table~\ref{rad_tab};
e.g. at RHIC we find 9.5~fm for pions, 10.2~fm for
kaons, 11.3~fm for protons, 11.6~fm for Lambda- and Sigma-Hyperons, 14.2~fm for
Cascades, but only 7.3~fm for the $\Omega^-$. The freeze-out of the 
$\Omega^-$ occurs rather close to the phase-boundary~\cite{dumitru99a}, due to
its very small hadronic interaction cross section. This observation
holds true for all three studied beam energies.
The respective {\em thickness} $\Delta r_{had}$ 
of the hadronic phase is reduced by a factor of 2 for the $\Omega^-$, compared
to that of the other baryon species.
This behavior could be responsible for the  experimentally observed
hadron-mass dependence of the inverse slopes 
of the $m_T$-spectra at SPS energies~\cite{thOm}.  
For the $\Omega^-$, the inverse slope remains practically unaffected by
the purely
hadronic stage of the reaction, due to its small interaction cross section,
while the flow of $p$'s and
$\Lambda$'s increases~\cite{dumitru99a} (see also section~\ref{flowsection}). 

\begin{figure}
\centerline{\epsfig{figure=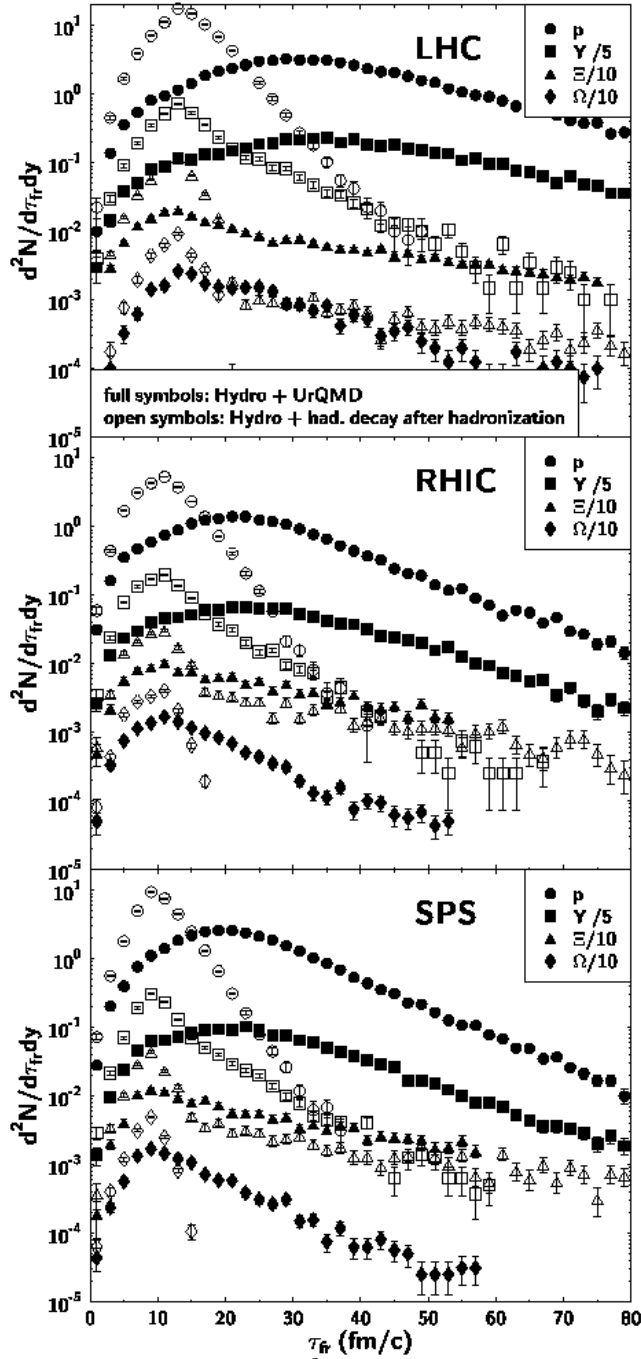,width=3.5in}}
\caption{\label{dndtf} Freeze-out time distributions
d$^2N/$d$\tau_{fr}$d$y$ of
$\pi$, $p$ and $\Omega^-$ for LHC (top), RHIC (middle) and SPS (bottom).
Open symbols denote the distributions for a pure hydrodynamical calculation 
up to hadronization with subsequent hadron resonance decays 
(but without hadronic reinteraction), whereas the full symbols show the 
full calculation with hadronic rescattering.}
\end{figure}

Figure~\ref{dndtf} shows the freeze-out time distributions 
d$^2N/$d$\tau_{fr}$d$y$ for
$\pi$, $p$ and $\Omega^-$ at LHC (top), RHIC (middle) and SPS (bottom).
Open symbols denote the distributions for a pure hydrodynamical calculation 
up to hadronization with subsequent hadron resonance decays 
(but without hadronic reinteraction), whereas the full symbols show the 
full calculation with hadronic rescattering. As we have already seen
previously in the transverse freeze-out radii, hadronic 
rescattering strongly modifies the shape of the distributions and significantly
increases the lifetime of the system. Table~\ref{tau_tab} lists the average
freeze-out times for $\pi$, $K$, $p$, $Y(=\Lambda+\Sigma^0)$, $\Xi$ and
$\Omega^-$ with and without hadronic rescattering.

One issue of great interest is the predicted significant increase of
the lifetime of the system from SPS to RHIC energies~\cite{RiGy96a},
being due to the time-delay caused by a first-order phase
transition~\cite{pratt}.
However, our model calculation (which does exhibit a first order
phase transition) shows no huge difference
in the freeze-out time distributions of
$\pi$, $p$, and $\Omega^-$ from SPS to RHIC energies (note, however, the
logarithmic scale). Origin of this prediction is that we
include many more states in the hadronic EoS, which speeds up hadronization
considerably~\cite{JSoll,DumRi,CRS}. Furthermore, decays of resonances
partly hide the remaining small increase of the hadronization time. 
Thus, the ``time-delay signal'' can not be expected to be well above
$\sim20-30\%$, and must be approached by a detailed excitation function.

Note that the
multi-strange $\Omega^-$ baryons freeze out far earlier than all other baryons,
as discussed already previously in the context of figure~\ref{dndrt}.
The {\em duration} of the hadronic reinteraction phase,
$\Delta \tau_{had} = \langle \tau_{fr}^{\rm Hydro+UrQMD}\rangle - 
\langle \tau_{fr}^{\rm Hydro + had. decays} \rangle $
remains nearly unchanged, e.g. at 5.9~fm/c for pions, 8.0~fm/c for kaons,
14.5~fm/c for protons, 15.4~fm/c for hyperons and 8.0~fm/c for the $\Omega^-$
between RHIC and SPS.

Note that the lifetime of the pre-hadronic stage in this approach 
is a factor of  $2-3$ longer than when employing the parton 
cascade model (PCM) \cite{geiger,bass99vni1} for the initial reaction stage. 
It will be interesting to check whether this is related to the 
first-order phase transition built into the EoS which is used here. 
The final transverse freeze-out radii and times (after hadronic rescattering),
however, are very similar in both approaches \cite{bass99vni1}.

\begin{figure}
\centerline{\epsfig{figure=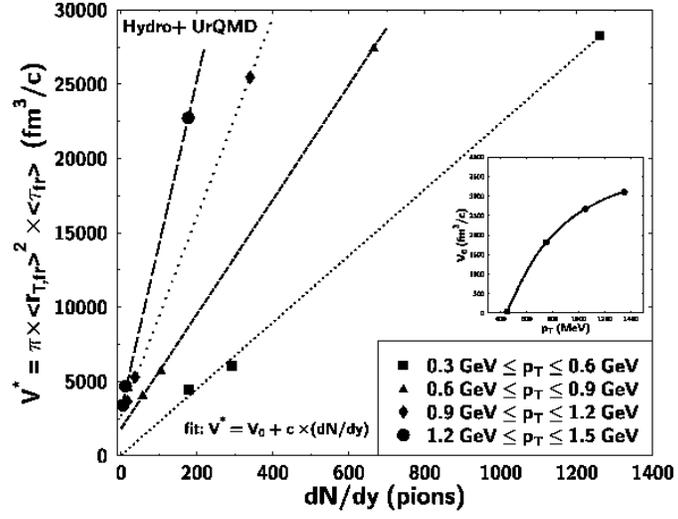,width=3.5in}}
\caption{\label{voldy2} Estimated freeze-out volume of pions as
a function of the pion rapidity density
for four different bins in transverse momentum.
High $p_T$-pions are only emitted from an outer 
{\em shell}, the radius of which increases with $p_T$. The inset shows
the volume of the ``hollow core'' as a function of $p_T$.}
\end{figure}

Figure~\ref{voldy2} shows the estimated freeze-out volume 
$V^*= \pi \langle r_{T,fr} \rangle^2 \tau_{fr}$ as
a function of the pion rapidity density d$N_\pi$/d$y$ 
for four different bins in transverse momentum. 
For all $p_T$-bins $V^*$ exhibits a nearly linear increase
with d$N_\pi$/d$y$. Thus, the freeze-out density of the
pions remains virtually constant over a large range of multiplicities
(or energies). We will see in the next section that this is due to the
fact that the chemical freeze-out of pions 
occurs rather shortly after hadronization of the QGP,
at all energies studied here. Since the local
density of pions on the hadronization hypersurface is similar in all
cases (because the temperature is almost the same), the density at
chemical freeze-out is, too.

We also observe that low-$p_T$ pions are basically emitted
from the entire volume, while at higher $p_T$ the pions seem only to be 
emitted from an outer {\em shell}, the radius of the hollow core
increasing with $p_T$. The inset of figure~\ref{voldy2} shows
the dependence of the non-emitting core-volume $V_0$ on the transverse
momentum of the pions. $V_0$ has been calculated by a linear
fit of $V^*$ to d$N_\pi$/d$y$: \mbox{$V^*=V_0 + c \, $(d$N_\pi$/d$y$)}.
The increase of $V_0$ with $p_T$ is a manifestation of the collective
flow effect; high-$p_T$ pions can not be emitted from the center,
$r_T\sim0$, since the collective velocity field vanishes there.

\subsection{Chemical freeze-out}
\label{ChemFO}

So far, we have only discussed the 
kinetic freeze-out of individual hadron species.
However, apart from the kinetic freeze-out, 
the chemical freeze-out of the system, which fixes the chemical
composition, is of interest, too.

The chemical freeze-out hypersurface of hadron species $i$ is in
principle defined as the surface $\sigma_{chem}^\mu$
separating the space-time region
where $\PD\cdot N_i=0$ from that where the number-current $N_i$ is
not conserved. Usually, the chemical freeze-out is defined modulo
hadronic resonance decays which are performed on $\sigma_{chem}^\mu$,
even for short-lived resonances like the $\rho$-meson or
$\Delta$-baryon. However, that definition is not very useful in the
present case, since most inelastic processes are actually modeled via
resonance excitation and subsequent decay, cf.\ section~\ref{UrQMD_section}.
Furthermore, as in the case of kinetic freeze-out studied above,
the microscopic transport model does not yield sharp hypersurfaces
(three-dimensional volumes) but rather freeze-out domains
(four-dimensional volumes). We shall therefore mainly discuss the evolution
of hadron multiplicities after hadronization, and their time-dependence.

\begin{figure}
\centerline{\epsfig{figure=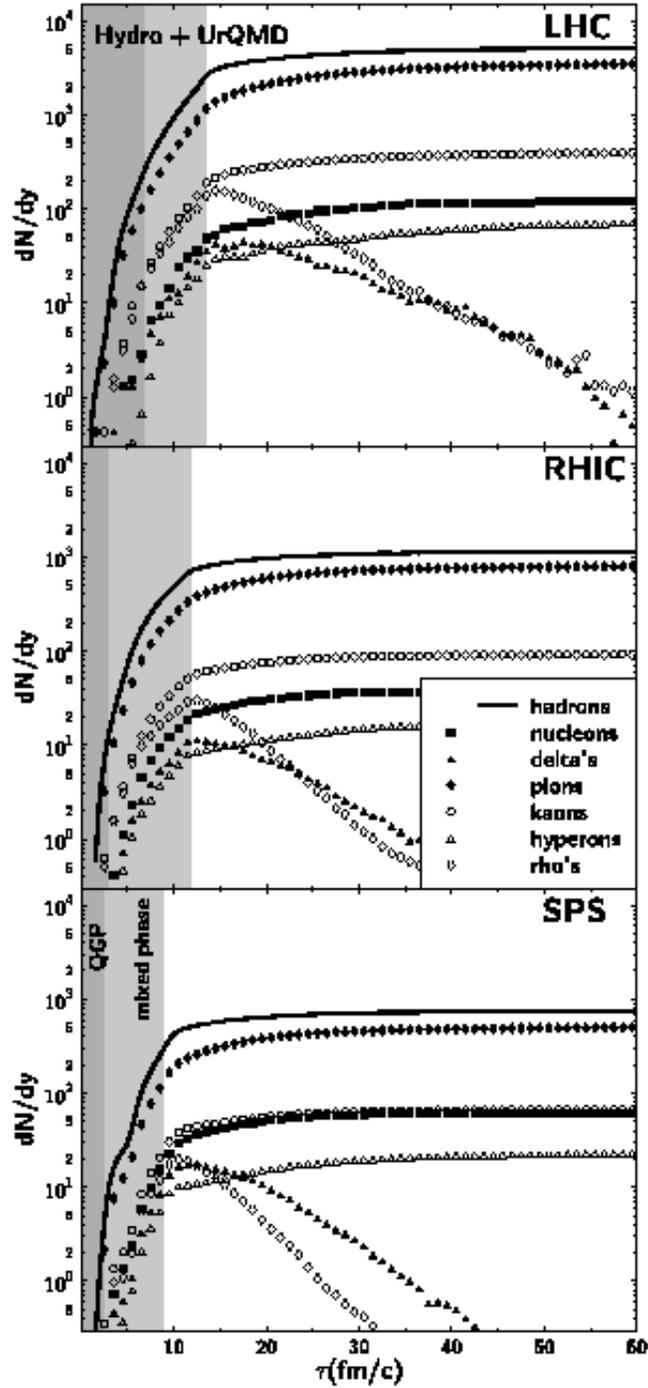,width=3.5in}}
\caption{\label{tevol_m} Time evolution of on-shell hadron
multiplicities (integrated over $r_T$) at LHC (top), RHIC (middle) and
SPS (bottom). 
The dark grey shaded area shows 
the duration of the QGP phase whereas the light grey shaded area depicts the
coexistence phase.}
\end{figure}

Figure~\ref{tevol_m} shows the time evolution of on-shell hadron
multiplicities for LHC (top), RHIC (middle) and SPS (bottom). 
The dark grey shaded area indicates the duration of the QGP phase,
whereas the light grey shaded area depicts the
mixed phase (both averaged over $r_T$; only hadrons that have already
``escaped'' from the mixed phase into the purely hadronic phase
are shown). Hadronic resonances are formed and are populated for a long time.
One can rather nicely observe the stronger transverse expansion as beam
energy increases: on $\tau={\rm const.}$ hypersurfaces
the resonance-decay ``tails'' get boosted to larger
$\tau$.
Due to those transversely boosted resonances the
hadron yields saturate only at rather large $\tau$,
approximately 25~fm/c at SPS and RHIC and about 40~fm/c at LHC.

By  comparing the final hadron yields resulting from
the hydrodynamical calculation 
(up to hadronization, including subsequent hadronic decays,
but no hadronic reinteractions) to that of 
the full calculation, which includes 
microscopic hadronic dynamics, we can quantify the changes of the 
hadrochemical content due to hadronic rescattering.

\begin{figure}
\centerline{\epsfig{figure=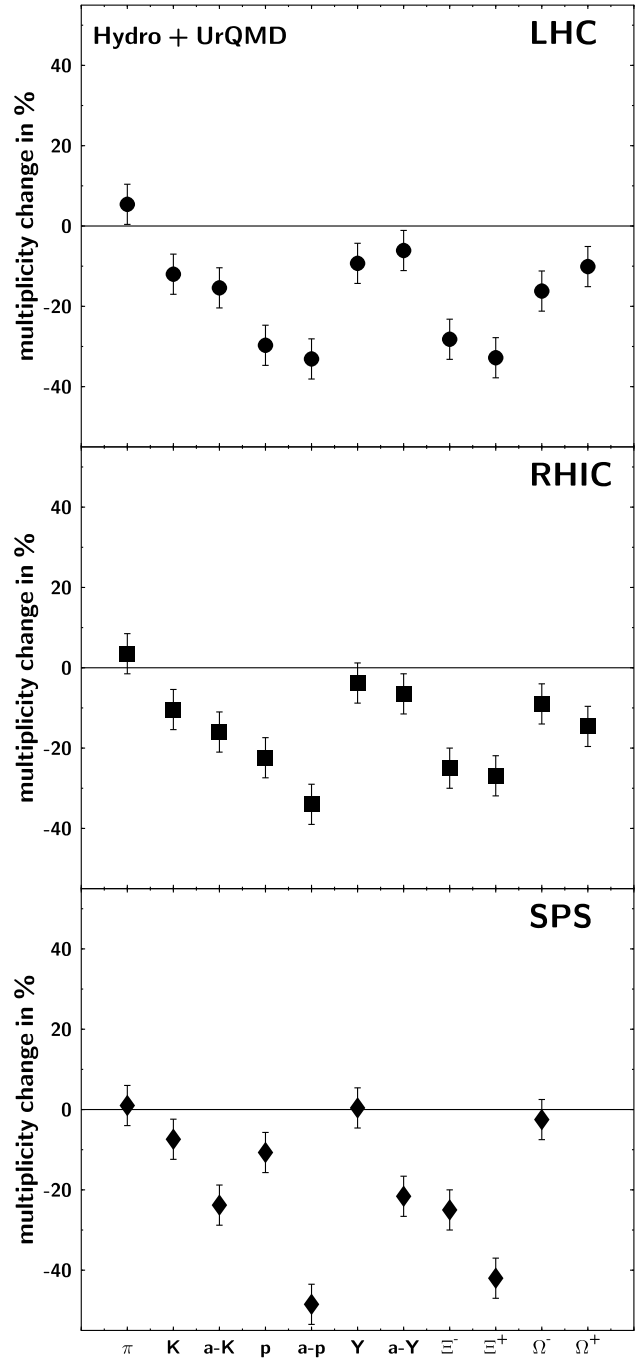,width=3.5in}}
\caption{\label{chemchange} Multiplicity change in \% due to hadronic
rescattering for various hadron species at SPS (bottom), RHIC (middle)
and LHC (top). The error-bars give an estimate of the systematic error.}
\end{figure}

Figure~\ref{chemchange} shows the relative change (in \%) of the
multiplicity for various hadron species for SPS (bottom), RHIC (middle)
and LHC (top). As to be expected, the state of rapid expansion
prevailing at hadronization does not allow chemical equilibrium
to hold down to much lower temperatures. The hadronic rescattering changes
the multiplicities by less than a factor of
two, cf.\ also~\cite{Stock:1999hm}. Thus, we have first evidence
that a QGP expanding and hadronizing as an ideal fluid produces a too
rapidly expanding background for a hadron-fluid with known
elementary cross-sections to maintain chemical equilibrium down to
much lower temperatures than $T_C$.

However, a closer look provides more insight into the
chemical composition. The changes are most pronounced 
at the SPS, were the baryon-antibaryon asymmetry is highest (since
the net-baryon density at mid-rapidity is highest). This manifests
e.g.\ in a reduction of the antiproton multiplicity by 40-50\% due
to baryon-antibaryon annihilation. $\bar{\Lambda}$ and $\bar{\Xi}$
are affected in similar fashion.

The baryon-antibaryon asymmetry decreases at higher beam energy, and at LHC
particle-antiparticle symmetry is restored for our initial
conditions.
The remaining small asymmetries (compare e.g.\ the $p$-$\overline{p}$,
$K$-$\overline{K}$, and $Y$-$\overline{Y}$ evolutions in Fig.~\ref{chemchange})
are due to fluctuations triggered by the finite number of particles,
which distort the ideal longitudinal boost-invariance present (by construction)
at hadronization.

Interestingly, the $\Omega^-$ multiplicity
decreases stronger towards higher beam energy. This is due to
the higher antibaryon density in the system, leading to more
$\Omega^-$ annihilations on antibaryons with subsequent redistribution
of the three strange quarks. (This process is modeled in UrQMD
as string excitation and subsequent fragmentation, cf.~\cite{uqmdref1}.)
Thus the hadronic phase becomes slightly more opaque
for the $\Omega^-$ with increasing beam-energy.

\begin{figure}
\centerline{\epsfig{figure=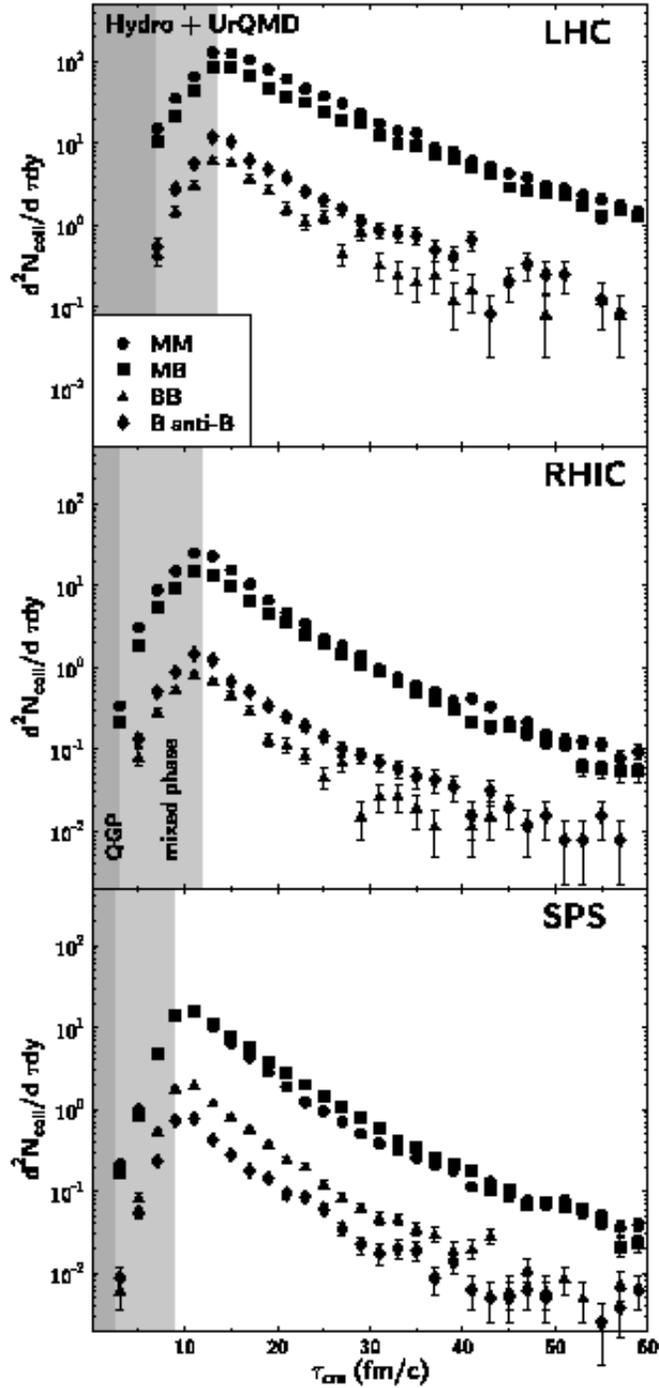,width=3.5in}}
\caption{\label{tevol_c} 
Hadron-hadron collision rates at LHC (top), RHIC (middle) and
SPS (bottom). 
The dark grey shaded area shows 
the duration of the QGP phase whereas the light grey shaded area depicts the
coexistence phase. }
\end{figure}

Collision rates offer another approach to determine the
duration of the hadronic phase, in particular $B$-$\overline{B}$ collisions
which almost always lead to annihilation.
Fig.~\ref{tevol_c} shows the time-evolution 
of the rates for  hadron-hadron collisions at LHC (top), RHIC (middle)
and SPS (bottom). 
Meson-meson (MM) and -- to a lesser extent -- meson-baryon (MB)
interactions dominate the dynamics in the hadronic phase at RHIC and
LHC. At the SPS meson-baryon and meson-meson interactions are equally
frequent. Note that while at the SPS baryon-baryon (BB) collisions
significantly outnumber baryon-antibaryon annihilations, the situation
at RHIC and LHC is reversed, where $B$-$\overline{B}$ annihilation is
far more frequent than BB collisions.
This is a consequence of the fact that the $B$-$\bar B$
annihilation cross sections at small relative momenta increase faster
then the total $B$-$B$ cross sections~\cite{uqmdref1}. 
In the case of (approximate) baryon-antibaryon symmetry, 
one therefore expects more $B$-$\bar B$
than $B$-$B$ interactions, as seen for RHIC and LHC energies.

Of course, all collision rates
reach their maxima at the end of the mixed phase, then decreasing
roughly according to a power-law.
After $\approx35$~fm/c, less than one hadron-hadron collision occurs
per unit of time and rapidity at SPS and RHIC energies;
due to the higher transverse $\gamma$-factor
the time is $\approx 60$~fm/c at the LHC.
At this stage the system is certainly
kinetically and chemically frozen-out.

\subsection{Transverse flow: 
         Emission of multi-strange baryons from the phase-boundary}
\label{flowsection}
In this section we analyze the transverse mass spectra at freeze-out,
and discuss their evolution from the hadronization hypersurface.
The results obtained for Pb+Pb collisions at CERN-SPS energy are in
reasonable agreement with the data obtained by the NA49-collaboration
\cite{NA49plmi} and by the WA97-collaboration
\cite{expOm}. For a comparison to that data we refer to~\cite{dumitru99a};
here, we focus on the model-results.

\begin{figure}[htp]
\centerline{\epsfig{figure=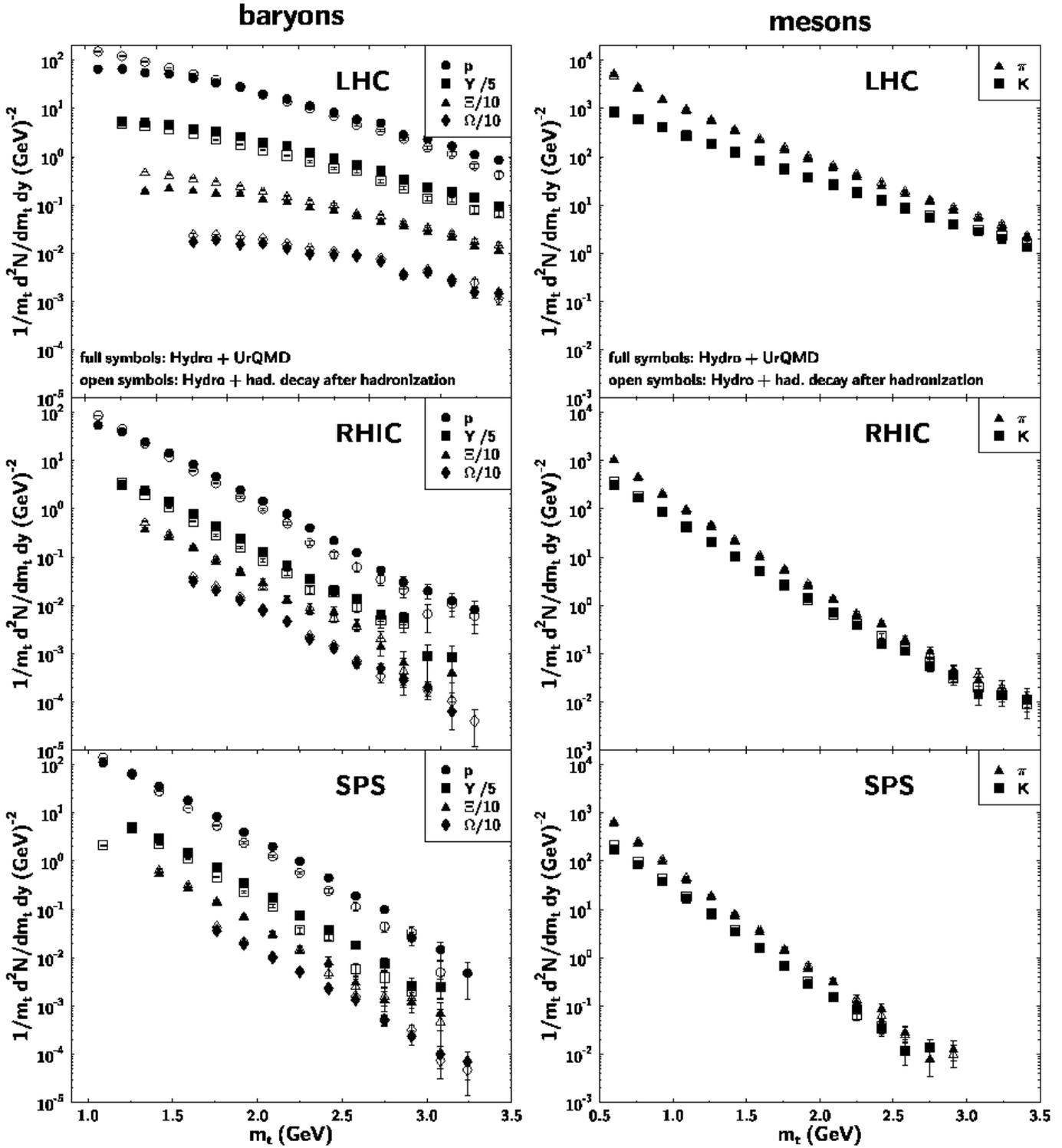,width=7.5in}}
\caption{Transverse mass spectra of $\pi$, $K$ (left column) and 
$N$, $\Lambda+\Sigma^0$,
$\Xi^0+\Xi^-$, and $\Omega^-$ (right column) 
at LHC (top), RHIC (middle) and SPS (bottom).
The open symbols denote the spectra on the hadronization hypersurface
whereas the full symbols show the calculation at freeze-out after hadronic
rescattering.}
\label{sps_mt}
\end{figure}

\begin{figure}[htp]
\centerline{\epsfig{figure=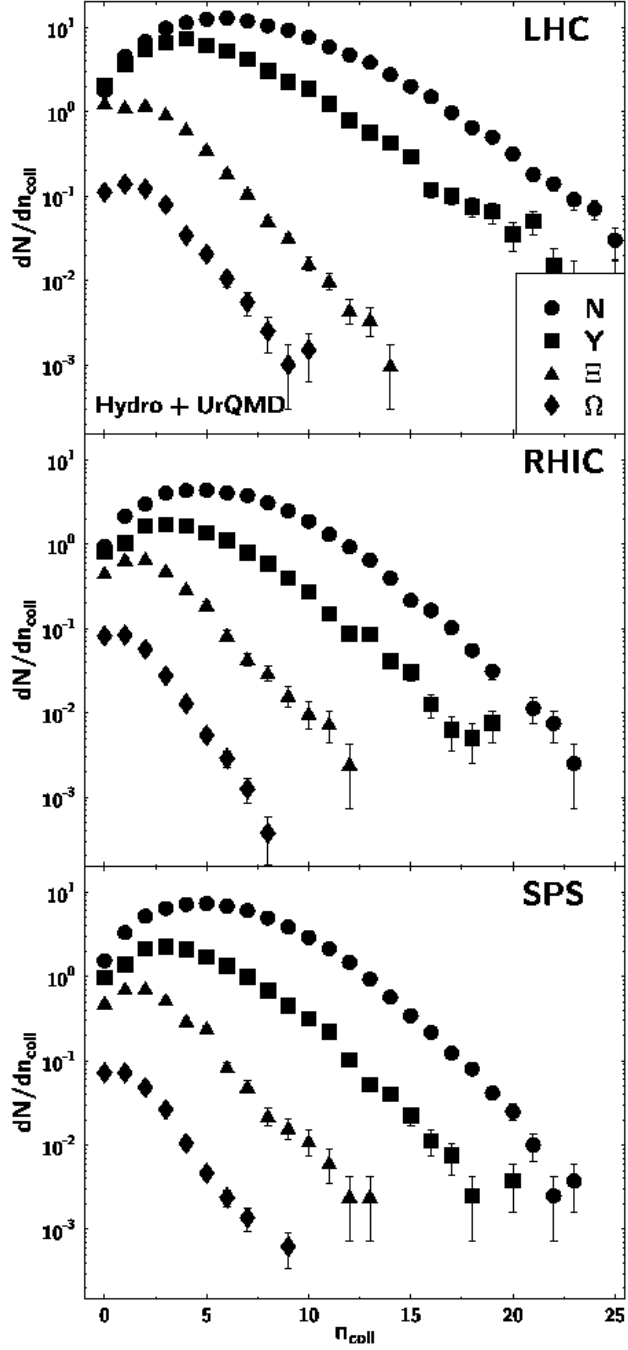,width=3.5in}}
\caption{Distribution of the number of interactions that the final-state
particles suffer after being hadronized; for LHC (top), RHIC (middle) and
SPS (bottom).}
\label{ncoll}
\end{figure}

Fig.~\ref{sps_mt} compares the $m_T$-spectra on the hadronization
hypersurface (open symbols), obtained from Eq.~(\ref{CFspectra}) (plus
strong resonance decays), with those at freeze-out (full symbols). 
One observes that
the transverse flow of $p$'s and $\Lambda$'s increases during
the hadronic stage, since those spectra flatten.
On the other hand, the spectra of $\Omega$'s and of $\Xi$'s with
$m_T\stackrel{>}{~}1.6$~GeV are practically
unaffected by the hadronic stage and closely resemble those on the phase
boundary. This is due to the fact that the scattering rates of $\Xi$ and
$\Omega$ in a pion-rich hadron gas are significantly smaller than those
of $N$'s and $\Lambda$'s~\cite{thOm,dumitru99a,hecke98a}. 
As shown in Fig.~\ref{ncoll}, on average
the baryons which finally emerge as $\Xi$'s and $\Omega$'s suffer
far less interactions than the final-state $p$'s and
$\Lambda$'s. Thus, within the model presented here,
these particles are basically emitted {\em directly from the phase boundary}
with very little further rescattering in the hadronic stage.
The hadron gas emerging from the hadronization of the QGP (in these
high-energy reactions) is almost ``transparent'' for the multiple
strange baryons. On the other hand, $p$'s and $\Lambda$'s on average suffer
several collisions with other hadrons before they freeze-out.
This behavior holds generally true for all three studied energy domains,
at the SPS, RHIC and LHC. 

These findings manifest themselves most strikingly in the
mass-dependence of the inverse slopes of the $m_T$ spectra.
A simple isentropic hydrodynamical expansion leads to broader
$m_T$ spectra
of heavier states, i.e.\ $\langle p_T\rangle$ or the inverse slope
$T^*$ increase with mass~\cite{Siem}. This observation agrees
with the inverse slopes of $\pi$, $K$, and $p$ measured for
central collisions of $Pb$ nuclei at a CM-energy of $17A$~GeV \cite{Tmsyst}.
However, it has also been found that the $\Xi$ and $\Omega$ baryons
do not follow this general trend~\cite{expOm,NA49plmi}.

\begin{figure}[htp]
\centerline{\epsfig{figure=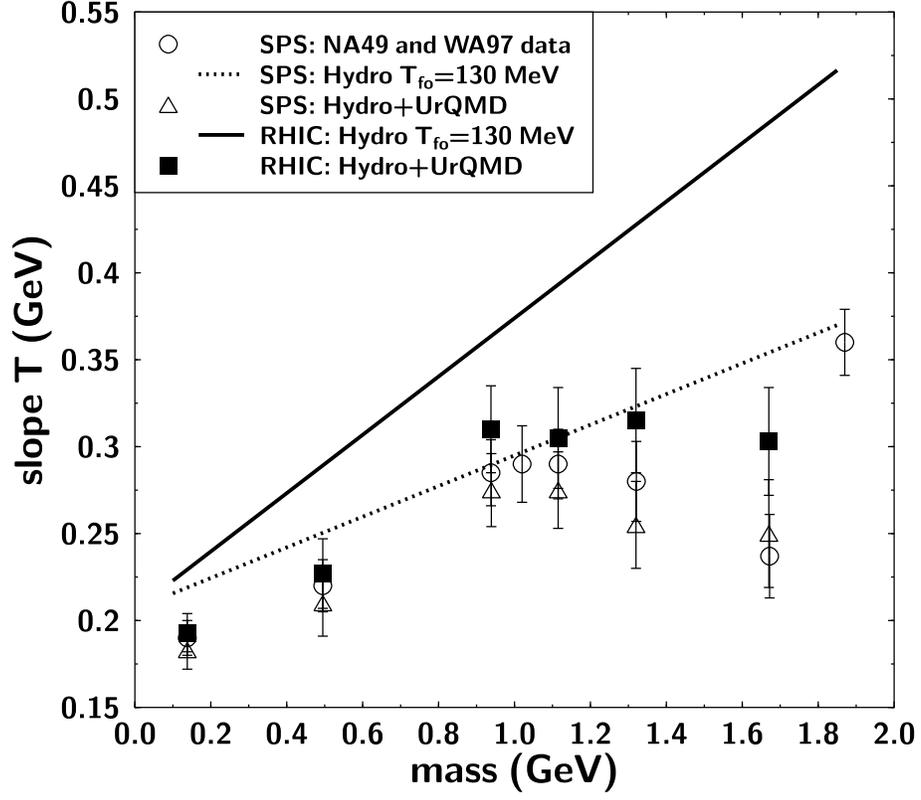,width=5in}}
\caption{Inverse slopes of the $m_T$-spectra of $\pi$, $K$, $p$,
$\Lambda+\Sigma^0$, $\Xi^0+\Xi^-$, and $\Omega^-$ at $y_{c.m.}=0$,
$m_T-m_i<1$~GeV.}
\label{slopes}
\end{figure}

Fig.~\ref{slopes} depicts the inverse slopes $T^*$ obtained from our model
by a fit of
$\d^3N_i/\d^2m_T\d y$ to $\exp(-m_T/T^*)$ in the range $m_T-m_i<1$~GeV.
The statistical error
of this fit is $\sim10\%$. Open symbols denote the SPS calculation and
data, whereas full symbols show the RHIC prediction. The  lines show
a purely hydrodynamical calculation~\cite{DumRi,dumitru99a}
with a freeze-out temperature of
$T_{fo}=130$~MeV for SPS (dotted  line) and RHIC (full line), respectively. 
The trend of the SPS data (open circles), 
namely the ``softer'' spectra of $\Xi$'s and
$\Omega$'s as compared to a linear $T^*(m)$ relation, is reproduced
reasonably well. As already mentioned, this is not the case for ``pure''
hydrodynamics with kinetic
freeze-out on a common hypersurface (e.g.\ the $T=130$~MeV isotherm), where
the stiffness of the spectra increases monotonically with mass, cf.\
Fig.~\ref{slopes} and also refs.~\cite{hung98a,KPjp}.
Resonance decays are not included in the hydrodynamic spectra
on the $T=130$~MeV isotherm.

When going from SPS to RHIC energy, the model discussed here generally
yields only a slight increase of the inverse slopes,
although the specific entropy is larger by a factor of 4-5~! 
The reason for this behavior is the first-order phase transition that
softens the transverse expansion considerably~\cite{soft}.
For our set of initial
conditions, the {\em average}
collective transverse flow velocity (at mid-rapidity)
on the hadronization hypersurface increases only from $\approx0.3$ (for
Pb+Pb at SPS) to $\approx0.35$ (for Au+Au at RHIC)~\cite{DumRi}.
(However, there are high-$v_T$ tails on the hadronization hypersurface
which get more pronounced at RHIC than at SPS.)
As can be seen from the present calculation, this is not counterbalanced
by increased rescattering in the purely hadronic stage --
compare to the
inverse slopes obtained from ``pure'' hydrodynamics with freeze-out on the
$T=130$~MeV isotherm!

\begin{figure}[htp]
\centerline{\epsfig{figure=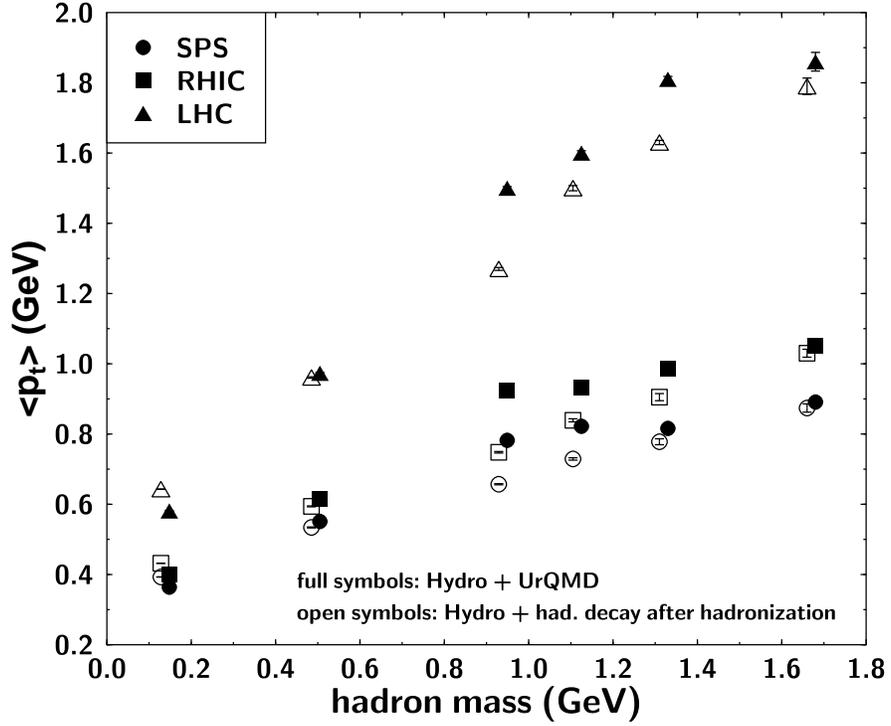,width=5in}}
\caption{Average transverse momentum $\langle p_T \rangle$ vs.\ mass
for $\pi$, $K$, $p$, $Y$, $\Xi$ and $\Omega$. 
The open symbols denote the spectra on the hadronization hypersurface
(including strong resonance decays),
the full symbols show the value at freeze-out (after hadronic
rescattering). For clarity, the symbols have been shifted by $\pm 10$~MeV.}
\label{ptm}
\end{figure}

The transverse flow at LHC beam energy is so strong that the $m_T$-spectra
can not be fitted any more by an exponential distribution.
We have therefore refrained to extract the slopes for the LHC calculation.
Instead, in figure~\ref{ptm} we show the mean transverse momenta
of the different hadron species as a function of their mass.
As in figure~\ref{sps_mt} we compare the $\langle p_T \rangle$ 
on the hadronization hypersurface (open symbols), 
obtained from Eq.~(\ref{CFspectra}) (plus strong resonance decays), 
with that at freeze-out (full symbols). Hadronic rescattering leads
to a transfer of transverse energy/momentum from  pions to heavier 
hadrons (the pions actually suffer a reduction of $\langle p_T \rangle$
in the hadronic phase)~\cite{kinDec}. This phenomenon has also been termed
the {\em pion-wind}~\cite{hung98a,oscar},
pushing heavier hadrons to higher $p_T$. 
Nucleons gain most transverse momentum, while
the $\Omega^-$ remains nearly unchanged due to its small interaction
cross section in the meson dominated
hadronic medium, as discussed earlier in this section. Those hadrons
are the best ``messengers'' of the early pre-hadronization evolution.

Furthermore,
one clearly observes the rather moderate increase
of $\langle p_T \rangle$ from SPS to RHIC energy, as
discussed already in Fig.~\ref{slopes}. In contrast, in our model
the collective
dynamics at the much higher CERN-LHC energy is dominated by the
stiff QGP, cf.\ also Fig.~\ref{h_hyper}, and the average
transverse momenta increase appreciably.

\section{Summary and outlook}

In summary, 
we have introduced a combined macroscopic/microscopic transport approach,
combining relativistic
hydrodynamics for the early deconfined stage of the reaction 
and the hadronization process
with a microscopic non-equilibrium model for the later hadronic
stage at which the hydrodynamic equilibrium assumptions are not valid anymore.
Within this approach we have self-consistently calculated the
freeze-out of the hadronic system,
accounting for the collective flow on the
hadronization hypersurface generated by the QGP expansion.

The reaction dynamics, hadronic freeze-out and transverse
flow in ultra-relativistic  heavy ion collisions
at SPS, RHIC and LHC have been discussed in detail. 
We find that the space-time domains of the freeze-out for the investigated
hadron species are actually four-dimensional, and differ drastically
between the individual hadrons species.

The {\em thickness} of the hadronic phase is found to be between 2~fm and
6~fm (at RHIC), depending on the respective hadron species. 
Its {\em lifetime} is between
5~fm/c and 13~fm/c, respectively. Freeze-out radii distributions have
similar widths for most hadron species, though the $\Omega^-$ 
is found to be emitted rather close to the phase boundary 
and shows the smallest freeze-out radii and times among all
baryon species.
The total lifetime of the system does not increase drastically when going
from SPS to RHIC energies. 

Our model-calculation shows that in high-energy nuclear collisions
the hadron multiplicities at midrapidity change by less than $40\%$
after hadronization, unlike e.g.\ in the early universe.
However, a closer look is warranted and reveals interesting
information. For example, {\em more} strange baryons ($\Lambda$,
$\Sigma$, $\Xi$, $\Omega$) are annihilated as the energy increases
because the anti-baryon density at hadronization increases.

Interactions within the hadron gas
increase the collective flow beyond that present at hadronization, and
reduce the temperature below the QCD phase transition temperature (we assume
$T_C=160$~MeV).
As an exception, we find that multiple strange baryons practically do not
rescatter within the hadron gas. Their $m_T$-spectra are therefore determined
by the conditions on the hadronization hypersurface, i.e.\ $T_C$ and
the collective flow created by the expansion preceding hadronization.
Their spectra therefore are less sensitive to the
confined phase, $T<T_C$, but are closely related to the EoS of the
QGP and the phase transition temperature $T_C$. 

Average transverse momenta and inverse slopes are predicted to increase
only moderately from SPS to RHIC, despite the significant increase
of the entropy to net baryon ratio. In this sense, the collective
evolution at RHIC energy is strongly characterized by the presence of
a well-mixed coexistence phase with small isentropic speed of sound.
It will be very interesting to see if this picture of hadronization of
bulk QCD matter, which is based on similar models for the QCD phase
transition in the much slower expanding early universe, agrees with the
data to be taken by the various experiments at BNL-RHIC.

Towards the much higher CERN-LHC energy, the evolution changes
appreciably. The pure QGP occupies a larger space-time volume than
the mixed phase. If the QGP-EoS at high energy density
is anywhere close to an ultrarelativistic
ideal gas with $p\sim\epsilon/3$, transverse expansion should be
much stronger than at RHIC and SPS. Consequently, the average transverse
momenta of the heavier hadrons increase by $60-80\%$ as compared
to RHIC energy.

We believe that the model presented here and in
refs.~\cite{bassdum99,dumitru99a} represents a step forward
towards the understanding and the description of the evolution
of a quark-gluon plasma, its hadronization, and the subsequent
freeze-out of the strong interactions. Nevertheless, it is
clear that many improvements are thinkable and necessary before
a really detailed comparison to
experimental data can be attempted. 

For example, corrections to
ideal fluid dynamics (before hadronization)
should be studied, at least within the Navier-Stokes approximation. 
In the present approach dissipative effects are only taken into
account after hadronization, where we expect them to be most
significant, particularly as freeze-out is approached.

The widely used Bag-model EoS can certainly be improved as well.
It is well known that it yields a substantially higher latent
heat than extracted from present lattice-QCD results. Thus, it
may over-pronounce the effects of a first-order QCD phase transition.
One may even try a cross-over transition to see whether that is
ruled out by experimental data or not.
Also, we have already commented on the fact that due to the huge
expansion rate in high-energy collisions
(which is not much smaller than strong interaction rates)
more radical scenarios like spinodal decomposition rather than an
adiabatic phase transition  should be examined as well.

To simplify the switch from hydrodynamics to the microscopic
transport model we assumed longitudinal boost-invariance and
azimuthal symmetry. The latter approximation, in particular,
disables us to study many up-to-date topics that will be addressed
by the experimental data, as e.g.\ anisotropies in the hadron
spectra (in non head-on collisions)
that may be sensitive to the QGP-EoS. A fully 3+1 dimensional
solution without symmetry assumptions is thus highly desirable.

One application not covered at all in the present studies is
correlated particle emission. Two-particle correlations may
for example allow to extract the volume of the emission region
in space-time.
The two-particle correlator probes rather soft
interactions and therefore contains information about the
freeze-out process (e.g.\ the thickness of the emission region,
background mean-fields etc.)~\cite{TPFO,pratt}.

Fluctuations in the rapidity and transverse momentum spectra
induced (or suppressed)
by the QCD phase transition are another highly
interesting topic. So far, we have studied only the evolution
on average, and have allowed only for fluctuations to develop
in the post-hadronization stage (which occur naturally in the
microscopic transport model).
However, the hadronization process could in principle induce
larger or other types of fluctuations, e.g.\ due to droplet
formation~\cite{igor}, spinodal decomposition with possible
DCC formation~\cite{HeisJack}, or due to the change of the order
of the phase transition in the vicinity of a second order
critical point~\cite{RajSh}. It would obviously be
highly desirable to know if they can survive the hadronic
interaction stage. Other fluctuations may be there from the
very beginning, e.g.\ those arising in the production process
of secondary hadrons, and one could study the evolution of
strangeness-rich rapidity bins~\cite{Spieles:1996is},
or of rapidity bins with
negative baryon-number, through the hadronization phase transition
until freeze-out.

There are many other questions that can not be listed here but
can be addressed within this model.
Work along those lines is in progress and will be reported in
forthcoming publications.

\acknowledgements
S.A.B.\ has been supported in part by the Alexander von Humboldt Foundation
through a Feodor Lynen Fellowship, and in part by DOE grant DE-FG02-96ER40945.
A.D.\ acknowledges support from the DOE Research Grant, Contract No.
De-FG-02-93ER-40764.
S.A.B.\ thanks Berndt M\"uller for many helpful and inspiring discussions.
A.D.\ thanks M.~Bleicher,
K.A.\ Bugaev, W.~Greiner, M.\ Gyulassy, D.H.\ Rischke, and H.\
St\"ocker for numerous inspiring discussions.\\
The computer-programs (implemented in FORTRAN 77)
with which the numerical calculations described in this
paper have been performed can be obtained free of charge (but copyright
protected) from the ``Open Standard Codes and Routines''
working group homepage, http://rhic.phys.columbia.edu/oscar.
The authors thank the UrQMD collaboration for the permission to use
the UrQMD transport model for the actual computations 
performed in this manuscript; and Dirk Rischke for the permission to use parts
of his RHLLE implementation of the fluid-dynamical equations, 
to solve the fluid-dynamical continuity equations.


\begin{table}
\begin{tabular}{cccccc}
\hline \hline
nucleon&delta&lambda&sigma&xi&omega\\  \hline \hline
$N_{938} $&$\Delta_{1232}$&$\Lambda_{1116}$&$\Sigma_{1192}$
&$\Xi_{1317}$&$\Omega_{1672}$\\
$N_{1440}$&$\Delta_{1600}$&$\Lambda_{1405}$&$\Sigma_{1385}$&$\Xi_{1530}$&\\
$N_{1520}$&$\Delta_{1620}$&$\Lambda_{1520}$&$\Sigma_{1660}$&$\Xi_{1690}$&\\
$N_{1535}$&$\Delta_{1700}$&$\Lambda_{1600}$&$\Sigma_{1670}$&$\Xi_{1820}$&\\
$N_{1650}$&$\Delta_{1900}$&$\Lambda_{1670}$&$\Sigma_{1775}$&$\Xi_{1950}$&\\
$N_{1675}$&$\Delta_{1905}$&$\Lambda_{1690}$&$\Sigma_{1790}$&$$&\\
$N_{1680}$&$\Delta_{1910}$&$\Lambda_{1800}$&$\Sigma_{1915}$&$$&\\
$N_{1700}$&$\Delta_{1920}$&$\Lambda_{1810}$&$\Sigma_{1940}$&$$&\\
$N_{1710}$&$\Delta_{1930}$&$\Lambda_{1820}$&$\Sigma_{2030}$&&\\
$N_{1720}$&$\Delta_{1950}$&$\Lambda_{1830}$&$$&&\\
$N_{1900}$&&$\Lambda_{2100}$&$$&&\\
$N_{1990}$&&$\Lambda_{2110}$&$$&& \\
$N_{2080}$ &&&&&\\
$N_{2190}$ &&&&&\\
$N_{2200}$ &&&&&\\
$N_{2250}$ &&&&&\\
\hline \hline
\end{tabular}
\caption{\label{bartab} Baryons and baryon-resonances treated in
the model. The corresponding
antibaryon states are included as well.}
\end{table}

\begin{table}
\begin{tabular}{cccccc}
\hline \hline 
$0^-$ & $1^-$ &$ 0^+$ &$ 1^+$ &$ 2^+$ & $(1^-)^*$\\ \hline \hline
 $\pi$ & $ \rho$ & $ a_0$ & $ a_1$ & $ a_2$ & $ \rho(1450)$ \\
 $K  $ &$   K^*$ & $ K_0^*$ & $ K_1^*$ & $ K_2^*$ &$ \rho(1700)$ \\
 $\eta$&$  \omega$& $ f_0 $&  $f_1$ & $ f_2 $ & $ \omega(1420)$ \\
 $\eta'$&  $\phi $&  $f_0^*$ & $ f_1'$& $ f_2'$ & $ \omega(1600)$ \\
\hline\hline
\end{tabular}
\caption{\label{mestab} Mesons and meson-resonances, sorted with
respect to spin and parity, treated in
the model.}
\end{table}

\begin{table}
\begin{tabular}{|c||rrr|rrr|rrr|} \hline
$\langle r_t \rangle$ (fm)
        & \multicolumn{3}{c|}{SPS} & \multicolumn{3}{c|}{RHIC} & 
\multicolumn{3}{c|}{LHC} \\\hline
species & d.p.b. & h.r. & h.t.    & d.p.b. & h.r. & h.t.     & d.p.b. & h.r. & 
h.t. \\\hline
$\pi$   & 6.9    & 8.4  & 1.5     & 7.8    & 9.5  & 1.7      & 12.3   & 16.8 & 
4.5  \\
$K$     & 6.2    & 8.4  & 2.2     & 7.1    & 10.2 & 3.1      & 11.4   & 18.1 & 
6.7  \\
$p$     & 4.7    & 9.1  & 4.4     & 5.4    & 11.3 & 5.9      & 8.7    & 22.2 & 
13.5 \\
$Y$& 5.1  & 9.5  & 4.4     & 5.8    & 11.6 & 5.8      & 9.6    
& 22.7 & 13.1\\ 
$\Xi$   & 8.3    & 12.1 & 3.8     & 9.4    & 14.2 & 4.8      & 8.7    & 22.1 & 
13.4 \\  
$\Omega^-$& 4.4  & 6.3  & 1.9     & 4.7    & 7.3  & 2.6      & 7.3    & 13.9 & 
6.6  \\
\end{tabular}
\caption{\label{rad_tab}
Mean transverse freeze-out radii $\langle r_T \rangle$ for different hadron 
species at SPS, RHIC and LHC. {\it d.p.b} denotes the  transverse freeze-out 
radii of the  hydrodynamical calculation up to hadronization, including 
subsequent hadronic decays, but no hadronic reinteractions, 
{\it h.r.} denotes the full Hydro+UrQMD calculation, including hadronic
rescattering and {\it h.t.} stands for an estimate of the
{\em thickness} $\Delta r_{had}$  of the hadronic phase:
$\Delta r_{had} = \langle r_{T,fr}^{\rm Hydro+UrQMD}\rangle -
\langle r_{T,fr}^{\rm Hydro + had. decays} \rangle$. }
\end{table}

\begin{table}
\begin{tabular}{|c||rrr|rrr|rrr|} \hline
$\langle \tau \rangle$ (fm/c)
        & \multicolumn{3}{c|}{SPS} & \multicolumn{3}{c|}{RHIC} & 
\multicolumn{3}{c|}{LHC} \\\hline
species & d.p.b. & h.r. & h.d.    & d.p.b. & h.r. & h.d.     & d.p.b. & h.r. & 
h.d. \\\hline
$\pi$   & 16.1   & 21.8 & 5.7     & 17.2   & 23.1 & 5.9      & 21.2   & 31.2 & 
10.0 \\
$K$     & 13.5   & 20.2 & 6.7     & 14.7   & 22.7 & 8.0      & 18.8   & 31.9 & 
13.1 \\
$p$     & 10.6   & 23.7 & 13.1    & 11.3   & 25.8 & 14.5     & 14.6   & 37.2 & 
22.6 \\
$Y$& 11.3 & 25.0 & 13.7    & 12.0   & 27.4 & 15.4     & 15.6   
& 39.0 & 23.4 \\ 
$\Xi$   & 19.9   & 31.0 & 11.1    & 20.4   & 32.2 & 11.8     & 14.1   & 36.2 & 
22.1 \\  
$\Omega^-$& 8.6  & 16.2 & 7.6     & 9.3    & 17.3 & 8.0      & 12.3   & 25.1 & 
12.8 \\
\end{tabular}
\caption{\label{tau_tab}
Mean freeze-out times for different hadron species at SPS, RHIC and LHC.
{\it d.p.b} denotes the freeze-out times of the  hydrodynamical calculation 
up to hadronization, including subsequent hadronic decays, but no hadronic 
reinteractions, {\it h.r.} denotes the full Hydro+UrQMD calculation, 
including hadronic rescattering and {\it h.d.} stands for an estimate of the
{\em duration} of the hadronic reinteraction phase,
$\Delta \tau_{had} = \langle \tau_{fr}^{\rm Hydro+UrQMD}\rangle - 
\langle \tau_{fr}^{\rm Hydro + had. decays} \rangle $.}
\end{table}

\newpage

\newpage

\newpage

\newpage

\newpage

\newpage

\newpage

\newpage

\newpage

\newpage


\newpage

\newpage

\newpage

\newpage

\end{document}